\documentclass[longauth]{aa}  

\usepackage{graphicx}
\usepackage{txfonts}
\usepackage{multicol}
\usepackage{multirow}
\usepackage{hyperref}
\usepackage{siunitx}
\usepackage{caption}
\usepackage{mathrsfs}
\usepackage{subcaption}
\usepackage{makecell}
\usepackage{float}
\usepackage{placeins}
\usepackage{chngcntr}

\DeclareMathOperator{\cmcube}{\si{\per\centi\meter\cubed}}
\renewcommand{\thesubfigure}{\roman{subfigure}}

\begin{document} 

   \title{Observations of Carbon Radio Recombination Lines with the NenuFAR telescope}
   \subtitle{I. Cassiopeia A and Cygnus A}

   \author{Lucie Cros\inst{1}
          \and
          Antoine Gusdorf\inst{1}\fnmsep\inst{2}
          \and
          Philippe Salom\'e\inst{2}
          \and 
          Sergiy Stepkin\inst{1}\fnmsep\inst{3}
          \and     
          Philippe Zarka\inst{4}
          \and
          Pedro Salas\inst{5}
          \and     
          Alan Loh\inst{4}
          \and
          Pierre Lesaffre\inst{1}
          \and
          Jonathan Freundlich\inst{6}
          \and
          Marta Alves
          \and
          Fran\c cois Boulanger\inst{1}
          \and
          Andrea Bracco\inst{1}\fnmsep\inst{7}
          \and
          St\'ephane Corbel\inst{8}
          \and
          Maryvonne Gerin\inst{2}
          \and
          Javier R. Goicoechea\inst{9}
          \and
          Isabelle Grenier\inst{8}
          \and
          Jean-Mathias Grie\ss meier\inst{10}\fnmsep\inst{11}
          \and
          Martin Houde\inst{12}
          \and
          Oleksandr Konovalenko\inst{3}
          \and
          Antoine Marchal\inst{13}
          \and
          Alexandre Marcowith\inst{14}
          \and
          Florent Mertens\inst{2}\fnmsep\inst{15}
          \and
          Fr\'ed\'erique Motte\inst{16}
          \and
          Michel Tagger\inst{10}
          \and
          Alexander Tielens\inst{17}
          \and
          Gilles Theureau\inst{10}\fnmsep\inst{11}
          \and
          Peter Tokarsky\inst{3}
          \and
          Oleg Ulyanov\inst{3}
          \and
          Vyacheslav Zakharenko\inst{3}
          }

   \institute{Laboratoire de Physique de l’\'Ecole Normale Sup\'erieure, ENS, Universit\'e PSL, CNRS, Sorbonne Universit\'e, Universit\'e Paris Cit\'e, F-75005, Paris, France \email{lucie.cros@phys.ens.fr}
         \and
            LUX, Observatoire de Paris, Universit\'e PSL, Sorbonne Universit\'e, 75014 Paris, France 
        \and
            Institute of Radio Astronomy NAS of Ukraine, 4, Mystetstv St., Kharkiv, 61002, Ukraine 
        \and 
            LIRA, Observatoire de Paris, Universit\'e PSL, Sorbonne Universit\'e, 75014 Paris, France 
        \and 
            Green Bank Observatory, Green Bank, WV 24944, USA 
        \and
            Universit\'e de Strasbourg, CNRS, Observatoire Astronomique de Strasbourg, UMR 7550, 67000 Strasbourg, France 
        \and
            INAF – Osservatorio Astrofisico di Arcetri, Largo E. Fermi 5, 50125 Firenze, Italy 
        \and 
            Universit\'e Paris Cit\'e and Universit\'e Paris Saclay, CEA, CNRS, AIM, 91190 Gif-sur-Yvette, France 
        \and
            Instituto de F\'isica Fundamental (CSIC). Calle Serrano 121-123, 28006, Madrid, Spain
        \and
            LPC2E, OSUC, Univ. Orl\'eans, CNRS, CNES, Observatoire de Paris, Universit\'e PSL, F-45071 Orl\'eans, France 
        \and
            ORN, Observatoire de Paris, Universit\'e PSL, Universit\'e d'Orl\'eans, CNRS, 18330 Nan\c cay, France
        \and 
            Department of Physics and Astronomy, The University of Western Ontario, 1151 Richmond Street, London, Ontario N6A 3K7, Canada
        \and
            Research School of Astronomy \& Astrophysics, Australian National University, Canberra ACT 2610 Australia
        \and
            Laboratoire Univers et Particules de Montpellier (LUPM) Universit\'e Montpellier, CNRS/IN2P3, CC72, place Eugène Bataillon, F-34095 Montpellier Cedex 5, France 
        \and
            Kapteyn Astronomical Institute, University of Groningen, PO Box 800, 9700 AV Groningen, The Netherlands
        \and
            Universit\'e Grenoble Alpes, CNRS, IPAG, 38000 Grenoble, France
        \and
            Leiden Observatory, Leiden University, Leiden, The Netherlands.   
            }

   \date{submitted 08/04/2025}

 
  \abstract
    {Carbon Radio Recombination Lines (CRRLs) at decametre wavelengths trace the diffuse phase of the interstellar medium (ISM) of the Galaxy. Their observation allows to measure physical parameters of this phase.}
    {We observed CRRLs with the recently commissioned New Extension in Nan\c cay Upgrading LOFAR (NenuFAR) telescope towards two of the brightest sources at low-frequency (10 to 85~MHz): Cassiopeia A and Cygnus A (hereafter Cas A and Cyg A respectively), to measure the density $n_{\rm e}$ and temperature $T_{\rm e}$ of electrons in line-of-sight clouds.}
    {We used NenuFAR's beamforming mode, and we integrated several tens of hours on each source. The nominal spectral resolution was 95.4~Hz. We developed a reduction pipeline, mostly to remove radio frequency interference (RFI) contamination and correct the baselines. We then performed a first fitting of the spectral lines observed in absorption, associated to line-of-sight clouds.}
    {Cas A is the brightest source in the sky at low frequencies and represents an appropriate test bench for this new telescope. On this source, we detected 398 C$\alpha$ lines between principal quantum numbers $n=426$ and $n=826$. Cyg A is also a bright source, however C$\alpha$ lines were fainter. We stacked the signal by groups of a few tens of lines to improve the quality of our fitting process. On both sources we reached significantly higher S/N and spectral resolution than the most recent detections by the LOw Frequency ARray (LOFAR). The variation of spectral linewidths with the electronic quantum number provides constraints on the physical properties of the clouds: $T_{\rm e}$, $n_{\rm e}$, as well as the temperature $T_0$ of the radiation field, the mean turbulent velocity $\varv_\mathrm{t}$ and the typical size of the cloud.}
    {Our final constraints differ from those inferred from LOFAR results, with, on average, $\sim$50\% lower $T_{\rm e}$, $\sim$35\% lower $n_{\rm e}$ and from 10 to 80\% higher $\varv_{\rm t}$. The NenuFAR observations sample a larger space volume than LOFAR's towards the same sources due to the differences in instrumental beamsizes, and the discrepancies highlight the sensitivity of low-frequency CRRLs as probes of the diffuse ISM, paving the way towards large area surveys of CRRLs in our Galaxy.}


\maketitle

\section{Introduction}
\label{sec:intro}
\paragraph{Radio recombination lines and the diffuse ISM.}
Our understanding of the life cycle of interstellar matter, from molecular clouds to ionised phases, encompassing star formation, is incomplete. Two key questions are related to (i) the processes of radiative and mechanical feedbacks exerted by young massive stars on their neutral interstellar environment, and (ii) the ability of the neutral interstellar gas to return to a molecular phase. Both questions will gain insight from the determination of the ionisation fraction, or equivalently the fractional abundance of free electrons. Indeed, although neutral in the sense that its main constituents (H and He) are in neutral forms, the neutral ISM can maintain a small ionisation fraction (from $\sim10^{-8}$ in its densest, molecular phases to $\sim10^{-1}$ in the warm neutral medium). Several processes contribute to this ionisation. Stellar FUV photons (energies between 6 eV and 13.6 eV, the latter value being the ionisation potential of hydrogen), can penetrate into neutral gas and ionise carbon and sulphur atoms (respectively of 11.26 and 10.36~eV ionisation potential). This is the case in photodissociation regions (PDRs), and more precisely in the unshielded surface layer of dense molecular clouds directly illuminated by nearby young massive stars (e.g., \citealt{Joblin18}, \citealt{Peeters24} and references therein). This is also the case more generally in the diffuse interstellar gas, widely permeated by the ambient Galactic stellar UV radiation \citep{Heays17}. Dynamical processes such as shocks can also partially ionise the gas (e.g., \citealt{Godard24}), whereas cosmic rays are the only possible source of ionisation in the densest parts of molecular clouds (\citealt{Dalgarno06}, \citealt{Padovani09}, \citealt{Padovani18}). Measuring the ionisation fraction in the ISM is paramount since beyond playing a key role in its chemistry, it controls its heating (e.g. through the photoelectric effect on interstellar dust grains), cooling (e.g. through the excitation of key species), and the excitation of key species (e.g. common molecular tracers with high dipole moments such as HCN and HCO$^+$). Until a few years ago, measuring the ionisation fraction had systematically been done by modelling the line emission from carefully selected species with astrochemical models \citep[see e.g.][for a review]{Bron21}.

Over the past 20 years, instrumental progress has resulted in the development of a new strategy to observationally constrain this parameter in the neutral interstellar regions, based on the observation of recombination lines, in particular from carbon, in various frequency ranges. Indeed, in presence of UV radiation, carbon is the main electron provider in these regions. Carbon atoms undergo a rapid cycle of ionisation followed by recombination with the free electrons present in the gas. During these recombinations, the species recombine in very excited electronic states and then cascade to lower energy states, emitting a large number of recombination lines. The shape of these recombination lines depends on the physical conditions - in particular the electron density and electron temperature - where the recombination occurs. They therefore constitute a probe of the ionisation fraction and physical properties of the diffuse ISM.

\paragraph{A review of recent RRLs observations.}
In the recent years, lines have been observed in the low frequency radio range ($<$~1~GHz) in absorption against bright Galactic radiosources: the carbon Radio Recombination Lines, or \lq CRRLs’. For instance, \citet{Stepkin07} observed series of carbon $\alpha$, $\beta$, $\gamma$ and $\delta$ recombination lines (hereafter \lq C$\alpha$’, \lq C$\beta$’, \lq C$\gamma$’, \lq C$\delta$’) in absorption towards Cassiopeia A (hereafter Cas A), respectively arising from principal quantum number $n+1, n+2, n+3, n+4 \rightarrow n$ de-excitations with the UTR-2Ukrainian radiotelescope around 26~MHz. Their detections of C$\delta$ lines with $n \in [625,1005]$ allowed them to estimate the electron temperature and density in line-of-sight clouds by fitting the evolution of the linewidth with the quantum number of the upper level of the transition. After \cite{Asgekar13} proved the feasibility to use LOFAR to study low-frequency C518$\alpha$ to C548$\alpha$ RRLs towards sources like Cas A, \citet{Oonk17} complemented this study of line-of-sight clouds towards this source with LOFAR observations in the 33 $-$ 78~MHz and 304 $-$ 386~MHz ranges. They detected series of hydrogen and carbon recombination lines in emission and absorption in three line-of-sight clouds, which led them to obtain better constrained measurements of the electron density $n_{\rm e}$ and temperature $T_{\rm e}$, to pinpoint the origin of the CRRLs in the CO-dark surface layers of molecular clouds where most of the carbon is ionised but the hydrogen is already molecular, and to estimate lower limits to the cosmic-ray ionisation rate. Simultaneously, \citet{Salas17} combined lower frequency (10 $-$ 33~MHz) observations of CRRLs with observations of the 158~$\mu$m [\ion{C}{II}] line to yield estimates of $n_{\rm e}$ and $T_{\rm e}$ in the same clouds. The values they found were consistent with those of \citet{Oonk17} and \citet{Stepkin07}. Finally, on this source, \citet{Salas18} and \citet{Chowdhury19} mapped CRRLs respectively with LOFAR (at 43, 54, 148 and 340~MHz) and the GMRT (in the 410 $-$ 450~MHz range). Their comparisons with other tracers of the ISM (\ion{H}{I}, OH, CO, \ion{C}{I}, \ion{C}{II}...) confirmed the utility of CRRLs as tracers of diffuse \ion{H}{I} and CO-dark gas halo’s around molecular clouds. The second brightest source in the low-frequency range, Cygnus A (hereafter Cyg A), was also observed with LOFAR by \cite{Oonk14}  in order to detect CRRLs. This study also constrained the electron density and temperature in one Galactic line-of-sight component. The smaller number of detected components could be explained by the higher Galactic latitude of the background source, $b = +5.76^\circ$ compared to $b = -2.13^\circ$  for Cas A. The modelling framework to extract physical parameters measurements from these Galactic observations of CRRLs, namely the level population problem of recombining carbon ions, was updated by \citet{Salgado17a, Salgado17b}. These authors calculated departure coefficients in the hydrogenic approximation including low-temperature dielectronic capture effects, and provided a study of CRRLs emission and absorption in a range of conditions in the diffuse, cold neutral medium (CNM). 

Beyond our Galaxy, a handful of studies report detections of RRLs in the low-frequency regime (below $\sim$1 GHz). First, \citet{Morabito14} discovered CRRLs in absorption in the M82 galaxy with a 8.5 sigma detection obtained by stacking 22 alpha transitions with n = 468 - 508 with the LOFAR telescope, although this result was only replicated with a lower SNR by \citet{Emig20}. They found that these lines were likely to be associated with cold atomic gas in the direction of the nucleus of M82, but their detection was not sufficient to measure physical parameters associated with the local medium. This shortcoming was remedied towards the 3C190 radio quasar at z$\sim$1.1946 by \citet{Emig19} with LOFAR observations. They demonstrated that H and/or C RRLs can be used to infer constraints on electron temperature and density in a line-of-sight dwarf galaxy about 80~Mpc from 3C190. \citet{Emig23} then reported the detection of hydrogen RRLs with the MeerKAT telescope around 1~GHz, from a $z\sim0.89$-galaxy that intercepts and lenses the PKS 1830-211 blazar. They were able to tie their detections to hydrogen, and to use them to measure the star formation rate within the lensing galaxy. Generally speaking, new perspectives in various directions are arising from recent RRL observations, in relation to observations of multi-ionised species \citep{Liu23}, heavier species than H, He and C (like sulphur, see \citealt{Goicoechea21a}), and thanks to the development of new reduction and interpretation techniques \citep{Emig20}. 

In the present study, we focused on the observation of C$\alpha$ recombination lines with the newly commissioned NenuFAR (New Extension in Nan\c cay Upgrading LOFAR) telescope towards the two brightest low-frequency radio sources: Cassiopeia A and Cygnus A, between 10 and 85~MHz. At these frequencies, CRRLs are detected in absorption in foreground interstellar clouds. In Sect.~\ref{sec:expset}, we briefly review the sources and the existing observations in similar ranges, and we present the NenuFAR instrument. In Sect.~\ref{sec:redpipe}, we present the reduction pipeline developed to recover up to 90\% of the 443 transitions within the frequency range. In Sect.~\ref{sec:results}, we present our fitting of the observed lines, highlighting our determinations of physical parameters such as the electron density and electron temperature. In Sect.~\ref{sec:discuss}, we compare our results to equivalent observations when possible. Sect.~\ref{sec:conc} contains our concluding remarks. 

\section{Experimental setup}
\label{sec:expset}

\subsection{The NenuFAR telescope}
\label{sub:tnt}

\begin{figure}[htbp]
    \centering
    \includegraphics[width=0.4\textwidth]{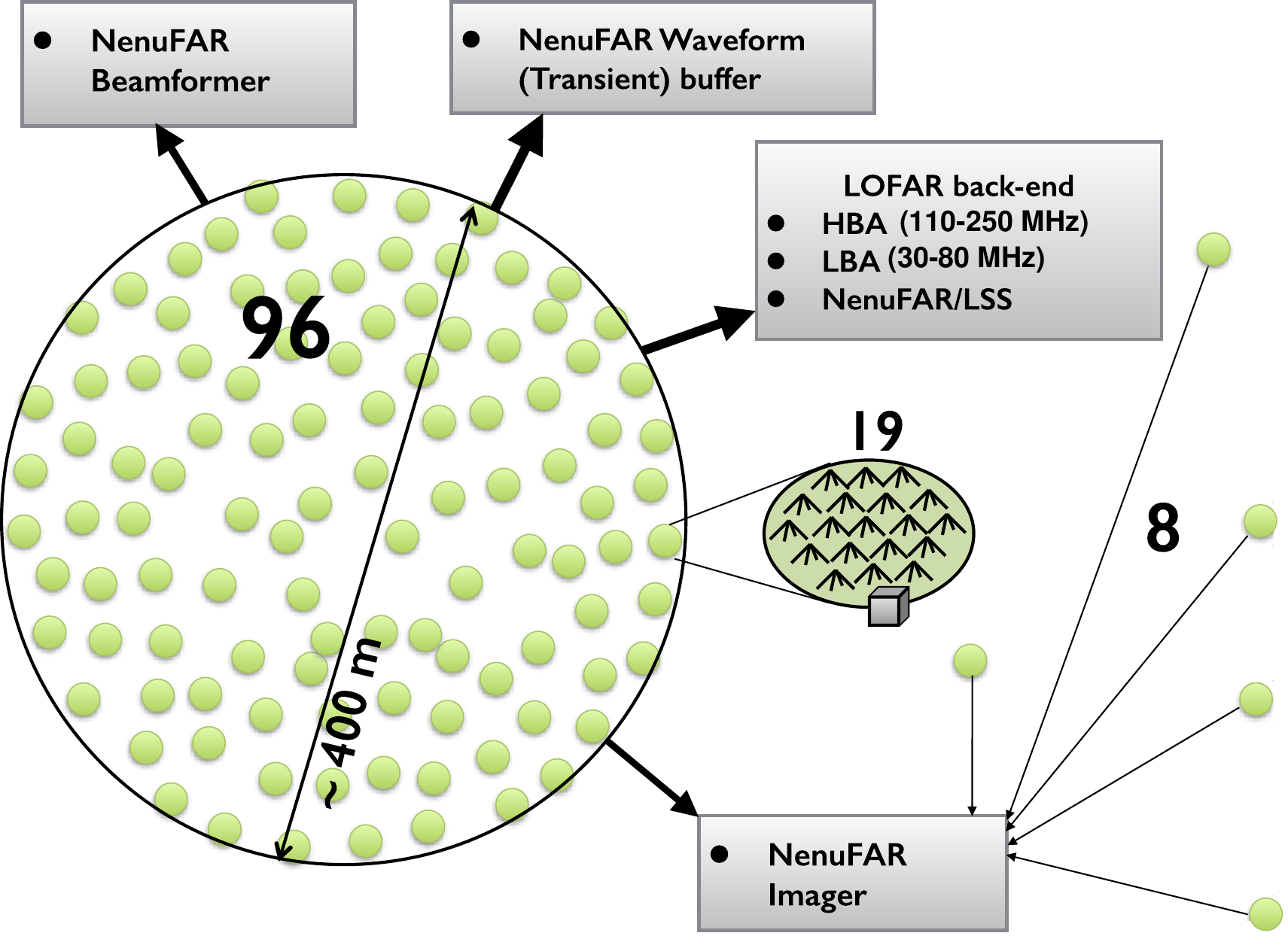}
    \includegraphics[width=0.25\textwidth]{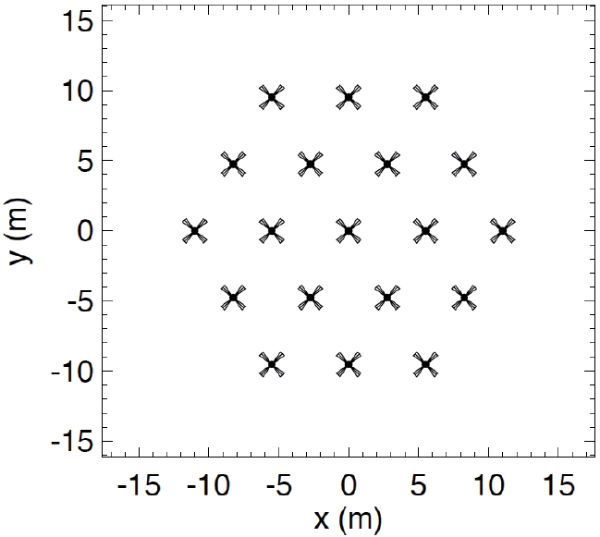}
   
    \caption{Configuration of the NenuFAR telescope. Top panel: layout of all 96 mini-arrays forming the NenuFAR interferometer. Bottom panel: Schematic layout of a mini-array showing the 19 inverted-V shape dual polarizations antennas. They are arranged in a 5.5 m triangular lattice forming a hexagon \citep[from][]{Girard23}. }
    \label{fig:nenufar-layout}
\end{figure}

The New Extension in Nançay Upgrading LOFAR (NenuFAR) \citep{Zarka20} telescope is located at the Nan\c cay Radioastronomy Observatory. NenuFAR is a low-frequency radio phased array and interferometer. We used it as a standalone phased array, although it will also be used as an extension of LOFAR. NenuFAR was designed as a pathfinder to the Square Kilometre Array (SKA\footnote{\href{https://www.skao.int/en/explore/precursors-pathfinders}{https://www.skao.int/en/explore/precursors-pathfinders}}). Its elementary receiving antennas consist of crossed inverted V-shaped dual-polarization dipoles. The antennas are organized in hexagonal tiles – called mini-arrays (MA) – that comprise each 19 such antennas. The MA beam of the hexagonal MA is analogically steerable across various directions in the sky. Due to the hexagonal symmetry, grating lobes are present in the individual MA primary beam \citep{Girard23}. To mitigate this, the MAs are rotated relative to each other by angles in multiples of 10 degrees, implying thus 6 non-redundant orientations. For now, NenuFAR is built at 85\% and features 80 MAs in its core\footnote{Upon completion, NenuFAR will feature 96 MAs within a 400~m diameter core and 8 remote MAs positioned at distances up to a few kilometers from the centre.}. Figure \ref{fig:nenufar-layout} illustrates the NenuFAR configuration\footnote{see \href{https://nenufar.obs-nancay.fr/en/astronomer/}{https://nenufar.obs-nancay.fr/en/astronomer/} for additional technical details about the NenuFAR telescope.}. The large dense core provides a high sensitivity, making it particularly effective for detecting faint low frequency radio recombination lines. All MAs are analogue phased using delay lines, and the entire array is digitally pointed quasi-continuously (at 10~s steps) towards the target.

Operating within the frequency range from the Earth’s ionospheric cut-off at $\sim$10~MHz to the radio FM band at $\sim$85~MHz, NenuFAR’s core provides varying angular resolution from 3.59$^\circ$ to 25$'$ for 80 MAs. The two linear polarizations outputs of each MA (NE-SW and NW-SE) are connected in parallel to multiple receivers, enabling operation in four distinct modes: standalone beam-former, standalone imager, waveform capture mode, and hopefully soon an upgraded LOFAR station mode that combines NenuFAR with the International LOFAR Telescope through the Nan\c cay LOFAR station (FR606). NenuFAR’s main backend LANewBa (for LOFAR-like Advanced New Backend) provides fluxes of complex waveform time series within 195.3125~kHz wide beamlets at a time resolution of 5.12~\si{\micro\second}. This flux can be numerically channelized down to $\delta f \simeq 95$~Hz at the expense of a lower time resolution of $dt \simeq 10$~ms. The sensitivity of NenuFAR presently ranges from $\sim$50~mJy at 15~MHz to $\sim$15~mJy at 85~MHz for a one-hour integration time with a frequency bandwidth of $\Delta f= 10$~MHz. At the nominal spectral resolution of 95.4~Hz, and for an on-source time of 2~h, the instrument achieves a sensitivity of $\sim$9~Jy at 15~MHz and $\sim$3~Jy at 85~MHz. These characteristics make NenuFAR one of the most sensitive radio telescopes operating below 85 MHz \citep{Zarka20}.

We used the NenuFAR telescope in beamforming mode, thanks to its UnDySPuTeD (for Unified Dynamic Spectrum Pulsar and Time Domain) receiver. It is fed by the main backend LANewBa, that is itself fed by the number of operating mini-arrays $\times$ 2 polarizations (NE and NW) in the core. Our observations began in 2019 and were performed within the Early Science Project 10 (ES10) of NenuFAR, which was converted into a so-called \lq Long Term' (LT) program, LT10 at the end of 2021. During the ES10 phase, 56 mini-arrays were operational, later evolving to 80 in the LT phase. LANewBa digitizes each of the \{56 to 80\} ×2 input signals at a sampling frequency of 200 MHz. The data streams are then channelized in \lq subbands' of 200 MHz/1024 = 195.3125 kHz bandwidth each. Beamforming then consists in coherently summing up the complex data streams from all operating MAs for a given subband, polarization and direction $(\theta, \phi)$ in the sky, after application of adequate phase shifts. A beamlet consists of one subband $\times$ two polarizations beamformed to one direction in the sky, meaning it is defined by a triplet $(\nu_{\rm c}, \theta, \phi)$ with $\nu_{\rm c}$ the central frequency of the subband. The raw data stream corresponding to one beamlet is the complex filtered waveform, that consists on average of 195312.5 pairs of complex X and Y signals values per second. LANewBa can compute and deliver 768 such beamlets at any given time, that correspond to an instantaneous usable bandwidth of 150 MHz. These 768 beamlets can be distributed in $(\nu_{\rm c}, \theta, \phi)$ as desired, for instance over two pointed directions simultaneously observed $(\theta_0, \phi_0)$ and $(\theta_1, \phi_1)$ with 75 MHz bandwidth (i.e. the full band of the instrument) to 768 beams paving the analogue beam at a single frequency $\nu_c$, through any intermediate configuration. In the rather exploratory phase of our program that we describe here, we used a configuration aimed at acquiring the full available bandwidth in only one direction of the sky at a given time, hence producing \lq only' 384 beamlets. 

The beamformed raw data from LANewBa were subsequently distributed via virtual local area network to several calculators, and especially to the 2 identical CPU/GPU \lq UnDySPuTeD' calculators, that further channelize the beamlet data down to 95~Hz resolution (2048 channels per beamlet), compute the 4 Stokes parameters from X and Y data, and perform time integration. Our observing setup consisted in using 384 beamlets over the 10.156 -- 84.961~MHz frequency range, totaling 786432 frequency channels with a resolution of $\sim$95.4~Hz. We set the time resolution to 84~ms with a Hamming-window apodisation. In this mode, the full width at half maximum (FWHM) of the observing beam was between 25.3$'$ (at $\sim$85~MHz) and 3.59$^\circ$ (at $\sim$10~MHz). The product of UnDySPuTeD is labelled \lq Level 0' (L0) data. At this stage the 384 beamlets were grouped in two \lq Lanes', labelled 0 and 1. The Lane 0 contains the first half of the spectra (from 10 to about 47~MHz) and the Lane 1 contains the second half of the spectra (from 47 to 85~MHz). The L0 data have a volume of the order of 260~GB/hr/lane. 

These data were subsequently reduced using an IDL software developed by Philippe Zarka and the NenuFAR team called \texttt{read\_nu\_spec} software, with aims to read, pre-process, reduce, and write \lq Level 1' (L1) data that have a volume reduced by a factor $\sim$100, stored in FITS format. This first reduction step performs a first pass of radio-frequency interference (RFI) mitigation, correction or whitening of bandpass and gain, and integration of the spectrum from the nominal 84~ms time resolution to a final resolution of 30~s. The \lq L1 data' output of \texttt{read\_nu\_spec} consists of a time-frequency table of Stokes $I$ values and a table containing statistical weights for each time-frequency integrated pixel determined by the percentage of data not flagged in the original time-frequency interval corresponding to that pixel.  
The time-frequency tables are subdivided in two subtables corresponding to the two \lq Lanes' of equal frequency width containing each 192 subbands of 2048 frequency channels each.  These level 1 data are the input of our reduction pipeline (see Fig.~\ref{fig:pipeline-cleaning}).

\subsection{Observed sources}
\label{sub:os}

We selected two radio sources to perform our experiment: Cassiopeia A and Cygnus A. They are the two brightest radio sources in the observed frequency range \citep{deGasperin20}, and they were previously observed by LOFAR LBA \citep{Oonk14, Oonk17, Salas17}. They hence optimize the chances of success of an absorption experiment, and represent an interesting benchmark for our new method of analysis using NenuFAR. The LOFAR observations of Cassiopeia A and Cygnus A were performed in imaging mode, and then the maps were cropped at the angular extension of the radio sources. Cassiopeia A (Cas A) is a supernova remnant and the brightest radio source in the northern sky, with 27104 Jy at 50 MHz \citep{deGasperin20}. CO \citep{Zhou18} and RRLs \citep{Oonk17,Salas17} observations revealed different velocity components along the line of sight in foreground clouds in the Perseus arm and the Orion spur. These clouds are detected at three different velocities: --47, --38 and 0~km~s$^{-1}$. Cygnus A (Cyg A) is a radio galaxy, and the second brightest radio source in the northern sky, with 22 146 Jy at 50 MHz \citep{deGasperin20}. In the foreground of Cygnus A, HI-21 cm \citep{Mebold75} and CRRLs \citep{Oonk14} show the presence of CNM at $\sim$3.5~km~s$^{-1}$ in the Orion spur.
Table~\ref{tab:source_info} provides the main characteristics and observation time of the sources. The largest angular size at 50 MHz of Cas A and Cyg A are respectively: $7.4'$ and $2.3'$. The beam of NenuFAR thus always encompasses the source.
\begin{table} 
    \centering
    \caption{{Characteristics of the radio sources.}}
    \begin{tabular}{|c|c|c|}
    \hline
         & Cassiopeia A & Cygnus A \\ \hline
         Object type & Supernova remnant & Radio galaxy \\ \hline
         Flux density & \multirow{2}{4em}{27~104~Jy} & \multirow{2}{4em}{{22}~146 Jy} \\ 
         at 50~MHz\tablefootmark{(a)}  & & \\ \hline
         Galactic longitude\tablefootmark{(b)}  & 111.7376$^\circ$ & 76.1899$^\circ$ \\
         Galactic latitude\tablefootmark{(b)}  & -2.1345$^\circ$ & +5.7553$^\circ$ \\ \hline
         On-source time & 71.5~h & 157.5~h \\ 
         \hline
    \end{tabular}
    \tablefoot{\tablefoottext{a}{see \cite{deGasperin20}}, \tablefoottext{b}{J2000}}
    \label{tab:source_info}
\end{table}

\section{Analysis pipeline}
\label{sec:redpipe}
\begin{figure}[htbp]
    \centering
    \includegraphics[height=22cm]{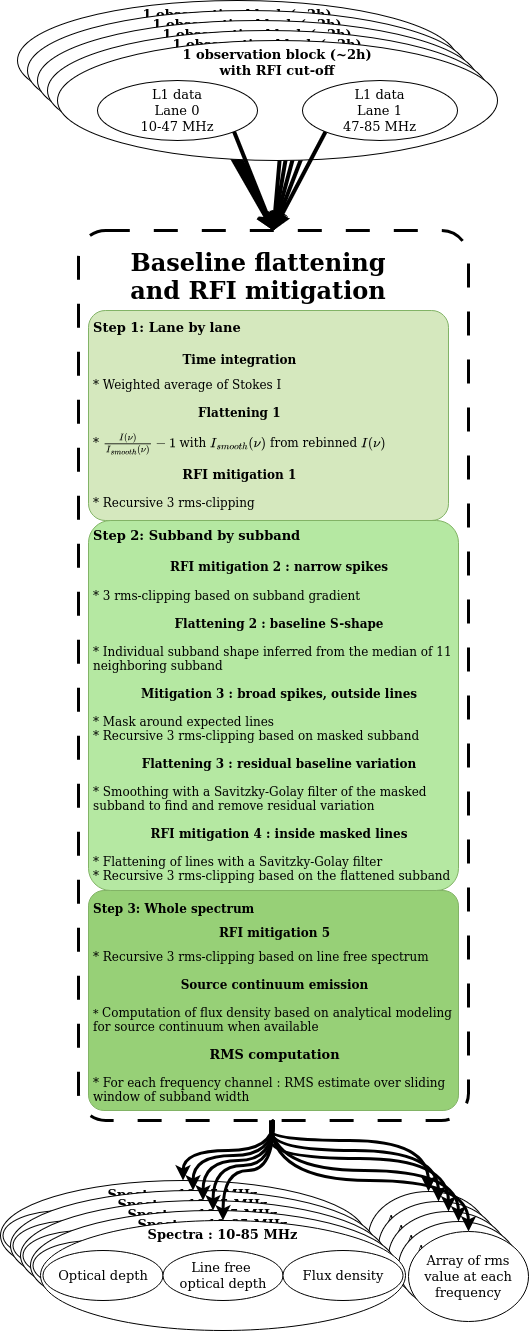}
    \caption{Functional diagram for the processing of L1 level data (see Sec.~\ref{sub:L1dp}). The algorithm is split in three reduction layers: lane-by-lane (upper part, 2a, Sec.~\ref{subsub:lbl}), subband-by-subband (middle part, 2b, Sec.~\ref{subsub:sbs}) and whole spectrum (lower part, 2c, Sec.~\ref{subsub:ws}). The processing consists in an iterative series of RFI mitigation and baseline flattening steps, and its application to real data is illustrated in Fig.~\ref{fig:reduc-lane-scale}.}
    \label{fig:pipeline-cleaning}
\end{figure}
From L1 data level, the analysis pipeline is divided in two major steps: a processing step aiming at providing a scientifically usable data product per observing block and per frequency channel, described in Sect.~\ref{sub:L1dp}, and a post-processing step consisting in a time integration and line stacking, described in Sects.~\ref{subsub:ta} and \ref{subsub:stacking}. 

\subsection{L1 data processing}
\label{sub:L1dp}

\begin{figure*}[htbp]
    \centering
    \includegraphics[width=0.9\textwidth]{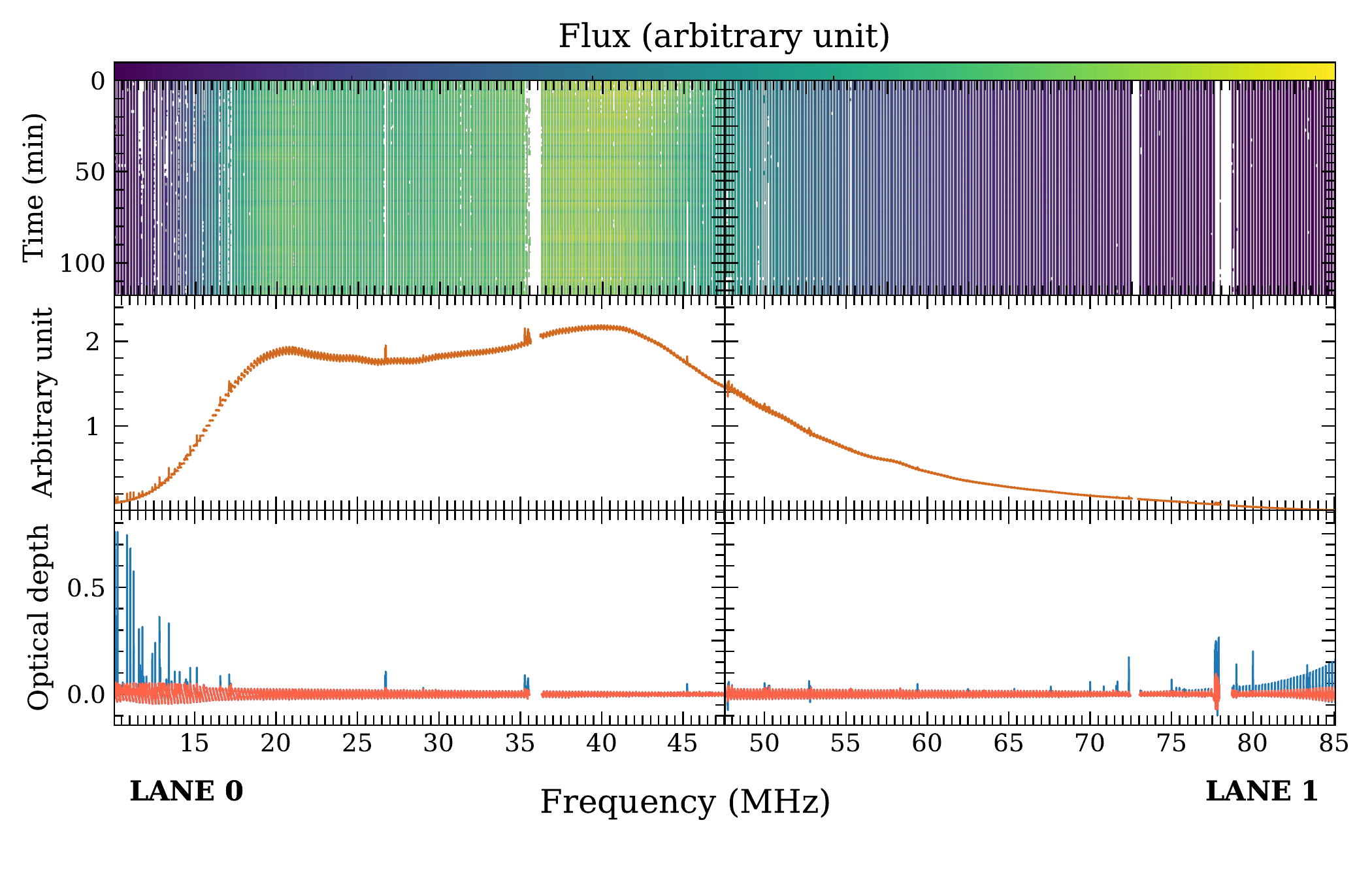}
    \caption{Description of the step 1 of the data processing algorithm (see top part of Fig.~\ref{fig:pipeline-cleaning}). The top panel represents the time-frequency table, $I(n_{\rm t}, n_{\nu})$, issued by the pre-processing algorithm \texttt{read\_nu\_spec}. The middle panel is the spectrum integrated over the two hours of observations, in arbitrary units. The bottom part shows the flattened spectrum before (blue) and after ({red}) 3~rms-clipping.}
    \label{fig:reduc-lane-scale}
\end{figure*}

The goal of the processing step is to remove RFI and flatten the continuum baseline, in order to produce one optical depth spectrum covering the whole 10 -- 85~MHz range per observing block. These goals are progressively achieved via (a) a first, large-scale, lane-by-lane correction described in Sect. \ref{subsub:lbl}, (b) a finer correction at subband scale described in Sect. \ref{subsub:sbs} and (c) the concatenation of all 384 subbands with a final mitigation over the whole spectrum described in Sect. \ref{subsub:ws}. A functional diagram is presented in Fig.~\ref{fig:pipeline-cleaning}. In the following, each step of the process is illustrated on the example of a 2-hours observing block acquired on the 20/09/2021 between 1 a.m. and 3 a.m. towards Cassiopeia A (see Figs.~\ref{fig:reduc-lane-scale}, \ref{fig:reduc-subband-scale}, \ref{fig:reduc-whole-scale}). 

\subsubsection{Lane by lane}
\label{subsub:lbl}

We first integrated the Stokes $I$ time-frequency L1 tables over the time axis weighted by the statistical weights determined by \texttt{read\_nu\_spec}, in order to obtain a single spectrum integrated over 2 hours (see top panels of Fig.~\ref{fig:pipeline-cleaning} for the functional diagram, and of Fig.~\ref{fig:reduc-lane-scale} for the data illustration). The integrated spectrum, noted $I$, shows variability on multiple scales: a wide broadband variability over the whole frequency range (see middle panel of Fig.~\ref{fig:reduc-lane-scale}) and a smaller narrowband variability at subband scale. To correct the wide variability, we determined empirically the global shape of the integrated spectrum by rebinning it to 192 points (one per subband) per lane. With a linear interpolation of the rebinned spectrum, we defined $I_{\rm smooth} (\nu)$, and compute the optical depth: $\tau (\nu) = 1 - \frac{I(\nu)}{I_{\rm smooth}(\nu)}$, at the nominal spectral resolution of our data. The linear approximation is sufficient at this stage, and potential deformations are dealt with in a subsequent step. Finally we applied a  3~rms-clipping on $\tau(\nu)$ to remove RFI, where rms is the root mean square of the spectrum. The rms is evaluated in a sliding window over the whole flattened lane. The size of the window is 2048 channels, which is the size of a subband. The bottom panel of Fig.~\ref{fig:reduc-lane-scale} shows the optical depth spectrum before and after the clipping.

\subsubsection{Subband by subband}
\label{subsub:sbs}

\begin{figure*}[htbp]
    \centering
    \includegraphics[width=\textwidth]{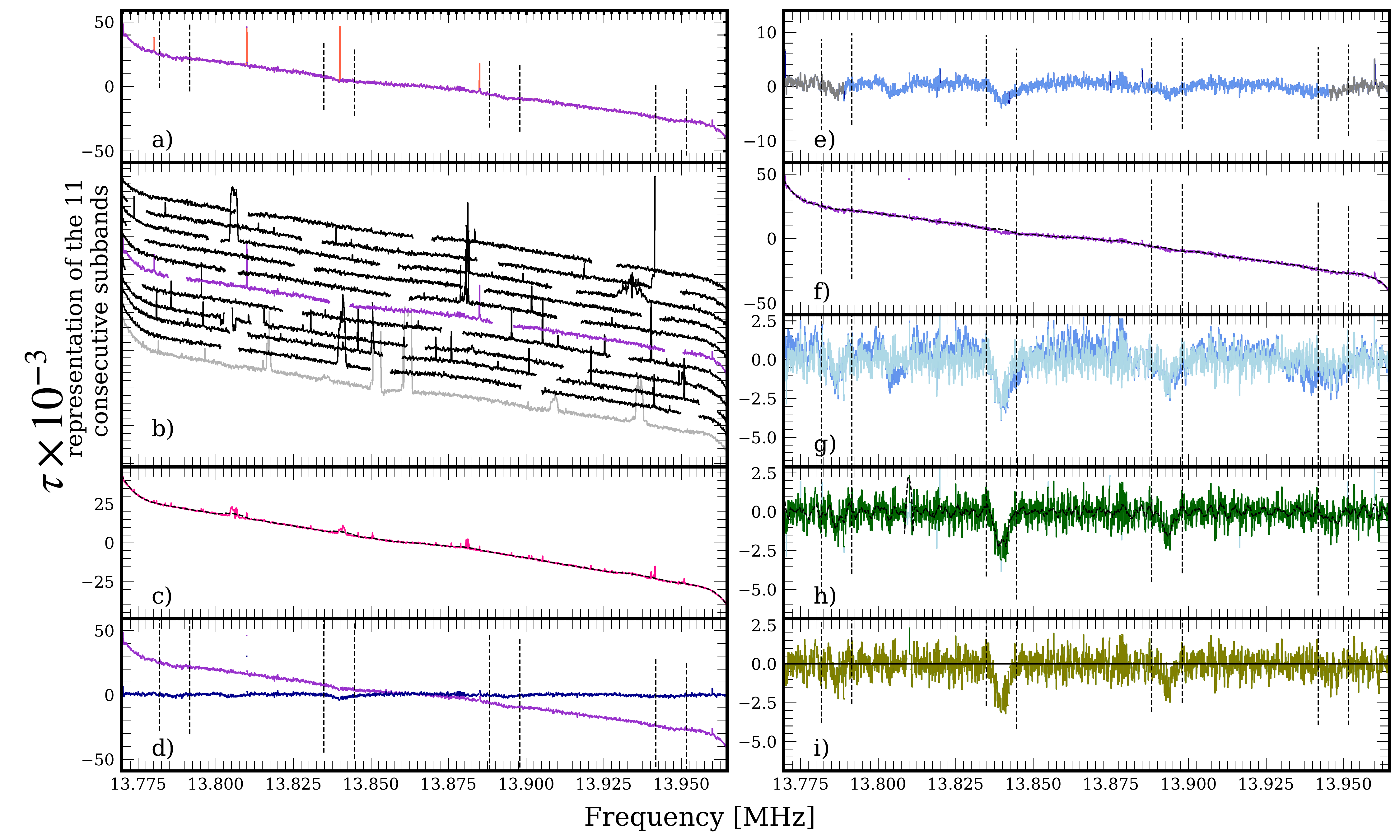}
    \caption{Description of step 2 of the data processing algorithm (see middle part of Fig.~\ref{fig:pipeline-cleaning}).  The data presented here corresponds to the 19$^\text{th}$ subband of the data set presented in Fig.~\ref{fig:reduc-lane-scale}. The black dotted lines represent the mask around the expected lines. The different colours represent the subband at each step of the processing (described in the middle part of Fig. \ref{fig:pipeline-cleaning}). a) The red (resp. purple) spectrum is before (resp. after) the \lq RFI mitigation 2' step. b) The purple spectrum is the subband being processed, the black spectra are the neighbouring subbands used to assess the general shape. The gray subband has been discarded. c) The pink line is the median of the black and purple subband in panel b), and the black dotted line is the median smoothed by a Savitzky-Golay filter. d) The purple (resp. navy blue) spectrum is before (resp. after) the \lq Flattening 2' step. e) The navy blue (resp. azure blue) spectrum is before (resp. after) the \lq RFI mitigation 3' step. In this case, RFI was removed at 13.92 and 13.96~MHz. g) The azure blue (resp. light blue) spectrum is before (resp. after) the \lq Flattening 3' step. The light blue (resp. green) spectrum is before (resp. after) the \lq RFI mitigation 4' step, and the black dotted line is the green spectrum smoothed by a Savitzky-Golay filter. i) The black line is the zero baseline and the olive green spectrum is the final flattened and clean subband.}
    \label{fig:reduc-subband-scale}
\end{figure*}

The next step consists in correcting for the narrowband variability and mitigating the RFI at subband scale. Indeed, the subband data from the NenuFAR beamformer generally presents an \lq S-shape' distortion, because of which some RFI have escaped the mitigation at lane scale. The various steps of the reduction applied to one subband are described in Fig.~\ref{fig:pipeline-cleaning}, and an illustration can be found in Fig.~\ref{fig:reduc-subband-scale}. We first mitigated RFI by targeting potential narrow spikes (over a few frequency channels) within each subband. To this aim, we performed a  3~rms-clipping on the gradient of the subband: anomalous spikes are thus identified and removed. This step is labelled \lq RFI mitigation 2' in Fig.~\ref{fig:pipeline-cleaning}, and the panel a) of Fig.~\ref{fig:reduc-subband-scale} shows the subband before and after the operation.

We then approximated the S-shape distortion of each subband by the median of 11 successive subbands ($\pm$ 5 subbands around the considered one, or $+k/-(10-k)$ if the subband is near the edges of the spectrum), median which is then smoothed with a Savitzky-Golay filter. In some cases, we had to discard anomalous subbands that presented a general shape that differed significantly from its neighbours. This operation is described in Appendix~\ref{sec:dasitcoags}. In order to preserve potential signal throughout all these operations, we applied a mask around the frequency of expected lines. The width of the mask around each predicted line has been determined empirically and depends on the source, but is typically $\sim$5.7~kHz ($\sim$60 frequency channels), that is 27~km~s$^{-1}$ at 85~MHz and 172~km~s$^{-1}$ at 10~MHz. Currently, this is a significant caveat of our analysis pipeline, in the sense that it makes human intervention necessary, as well as the prior knowledge of the velocities at which the lines are expected. Fixing this caveat is work in progress. The approximation of the S-shape distortion is labelled \lq Flattening 2' in the diagram Fig.~\ref{fig:pipeline-cleaning}, and is illustrated in the panels b) to d) of Fig.~\ref{fig:reduc-subband-scale}.

We subsequently applied a new  3~rms-clipping on the flattened subbands, in order to mitigate broad RFIs outside the masks (see the step labelled \lq RFI mitigation 3' in Fig.~\ref{fig:pipeline-cleaning}, and panel e) of Fig.~\ref{fig:reduc-subband-scale}). At this stage, the lines were still masked, and residual distortion of the baseline within the considered subband still subsisted, especially on the edges. We modelled this residual variation by subtracting a smooth baseline computed from a Savitzky-Golay filter applied on the masked subband (see the step labelled \lq Flattening 3' in Fig.~\ref{fig:pipeline-cleaning}, and panels f) and g) of Fig.~\ref{fig:reduc-subband-scale}). 

This enables us to remove RFI inside the masked parts of the subband. To do so, we temporarily flattened the inside of the mask using a Savitzky-Golay filter, and applied a  3~rms-clipping (see the step labelled \lq RFI mitigation 4' in Fig.~\ref{fig:pipeline-cleaning}, and panel h) of Fig.~\ref{fig:reduc-subband-scale}). The parameters of the filter are the same throughout the spectrum. We eventually obtained a flat subband, freed from most of its original RFIs (see panel i) Fig.~\ref{fig:reduc-subband-scale}{, which represents the final spectrum}). 

\subsubsection{Whole spectrum}
\label{subsub:ws}

\begin{figure*} 
    \centering
    \includegraphics[width=0.8\textwidth]{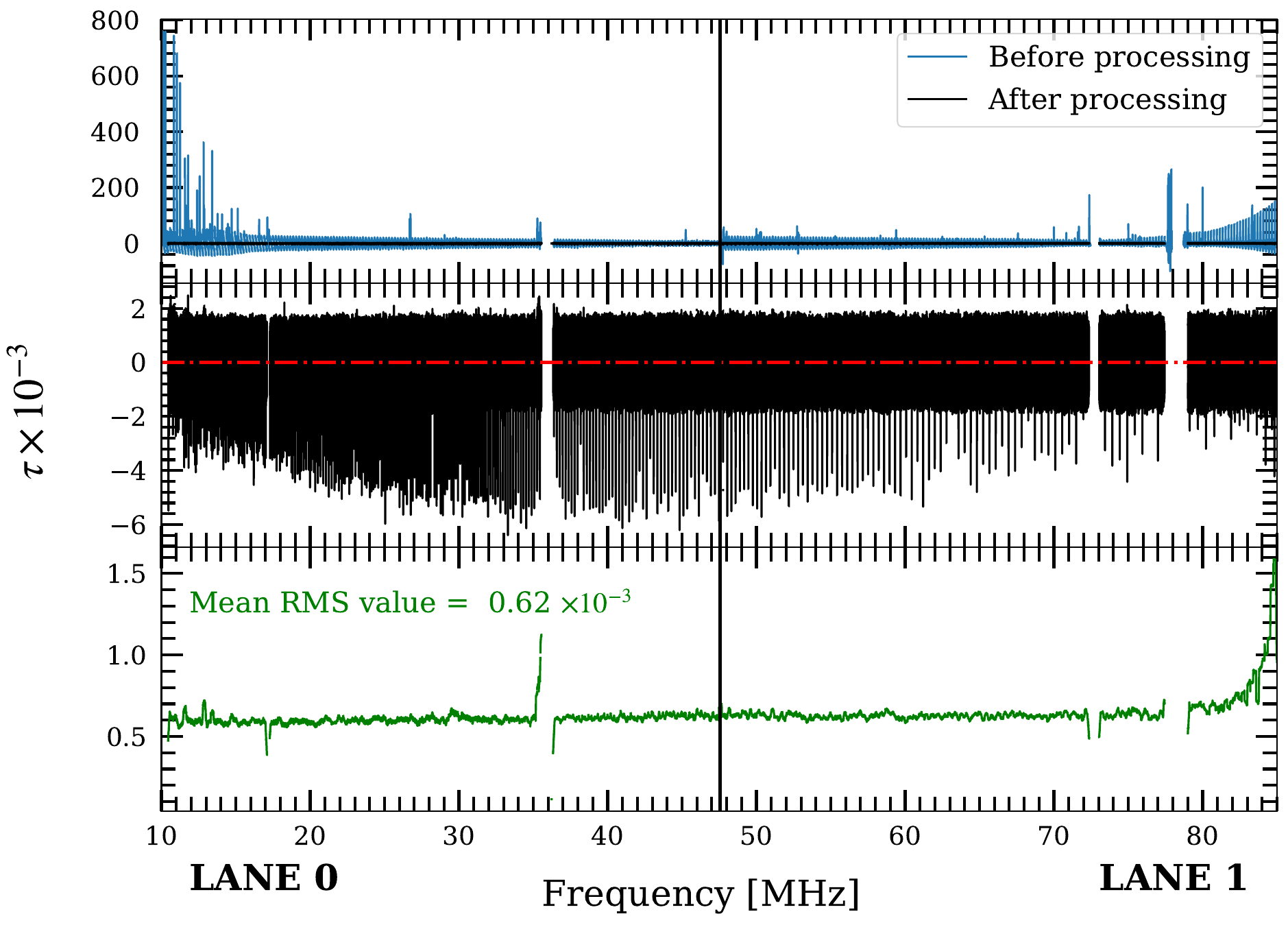}
    \caption{{Description of the step 3 of data processing (see bottom panel of Fig.\ref{fig:pipeline-cleaning})}. The black vertical line shows the separation between Lane 0 and Lane 1. The top panel compares the flat spectrum before (blue) and after (black) the mitigation process. The middle panel is a zoom on the black spectrum. The red dashed-dotted line shows the zero baseline. The vertical spikes are detections of C$\alpha$ RRLs. The bottom panel shows the value of the rms computed in a sliding window along the spectrum. The holes in the spectrum (at 35~MHz, 73~MHz and 79~MHz) are flagged regions contaminated by unmanageable RFI pollution during this observing block.}
    \label{fig:reduc-whole-scale}
\end{figure*}

Finally, we concatenated all 384 subbands to build a global spectrum spanning from 10 to 85~MHz, to which we applied a final  3~rms-clipping procedure. The goal was to remove remaining spikes located on the edges of the subbands, as well as to mitigate remaining RFIs according to the global noise level of the spectrum. We computed the rms values over a sliding, 2048 channels-wide window along the whole spectrum. These values are used as a metric for the \lq quality' of each frequency channel in the averaging and stacking procedures. This final step of the data processing is illustrated in Fig. \ref{fig:reduc-whole-scale}.

\subsection{Time averaging}
\label{subsub:ta}
\begin{figure*} 
    \begin{subfigure}[b]{0.6\textwidth}
        \includegraphics[width=\textwidth]{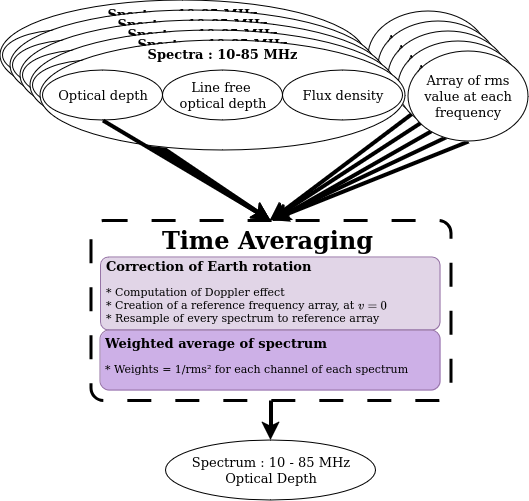}
        \caption{Time averaging}
        \label{fig:time-averaging}
    \end{subfigure}
    ~
    \begin{subfigure}[b]{0.4\textwidth}
        \includegraphics[width=\textwidth]{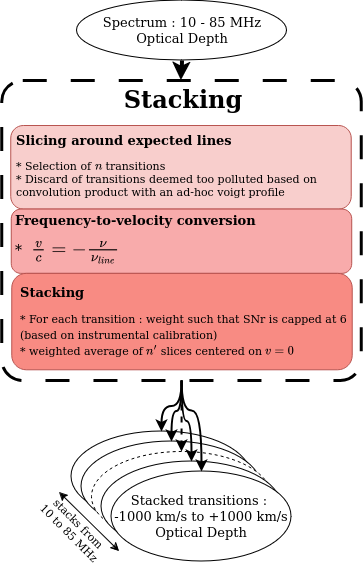}
        \caption{Stacking}
        \label{fig:stacking}
    \end{subfigure}
    \caption{Functional diagrams of the post-processing of the data. The left part of the figure describes the time averaging process  described in Sect.~\ref{subsub:ta}. The right part illustrates the stacking process described in Sect.~\ref{subsub:stacking}.}
\end{figure*}

The reduction pipeline is applied on each 2 hours observation block. The resulting spectra are then averaged together (see Fig.~\ref{fig:time-averaging}). We note that the RFI depend greatly on human activity, and thus on the moment when the observations were performed. Two criteria were used to assess the quality of an observation block before time averaging: the percentage of frequency channels flagged out as RFI, and the rms of the spectrum. The former is mainly influenced by whether the observation was performed during the day or during the night, with up to an average of 80\% of the spectrum flagged for observation around noon, and above 30\% between 5 am and 7 pm. Heavy flagging of daytime spectra allows the rms on the remaining frequency channels to be stable throughout the day. Neither of these criteria are impacted by the day, the month, or the year of the observation.
For Cygnus A, the mean rms value of the optical depth noise goes from ~$3\times 10^{-4}$ to over ~$8\times 10^{-4}$ depending on the moment the observations were performed. For Cassiopeia A, the mean rms value of the optical depth goes from ~$4\times 10^{-4}$ to about ~$7\times 10^{-4}$.

A blind averaging procedure would allow narrowband sporadic RFI to propagate throughout all the spectra. As a result, we used the rms of the spectra computed on a sliding window of length = 2048 frequency channels, as a way of assessing the quality of a spectrum. We compute the weight $k_{\nu,\mathrm{S}}$ for each frequency channel $\nu$ of each spectrum $S$ as:
\begin{equation}
    k_{\nu,\mathrm{S}} = \frac{1}{\sigma_{\nu,\mathrm{S}}^2}
\end{equation}
where $\sigma_{\nu,\mathrm{S}}$ is the rms of the spectrum $S$ at the frequency $\nu$. 
To account for the Doppler effect due to the rotation of the Earth with respect to the kinematical local standard of rest, we realigned each spectrum on a reference frequency grid, before averaging\footnote{We do not correct the Doppler shift within a 2-hours observation, because the Doppler shift between the first and last spectra, although it varies throughout the year, is at most of $\sim$100 m/s, thus always below the spectral resolution of NenuFAR ($\sim$2.6 km/s at 10 MHz, $\sim$335 m/s at 85 MHz).}. The weighted average of the spectra is computed to create a single spectrum with an increased S/N for further analysis. 

\begin{figure*} 
\centering
        \includegraphics[width=0.8\textwidth]{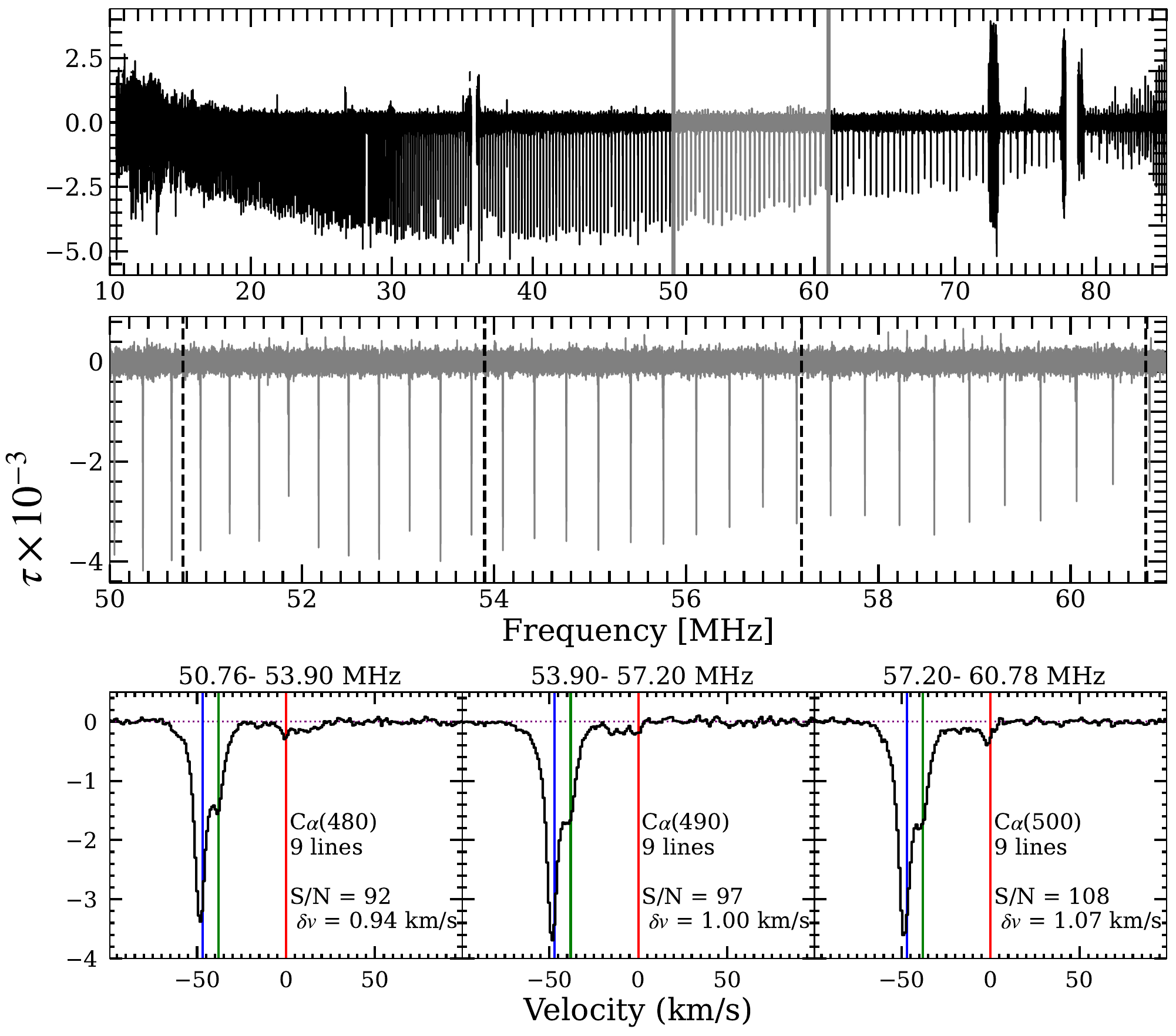}
    \caption{Post-processing of the data, on the example of Cassiopeia A. The top panel represents the whole spectrum averaged over all observing blocks. The gray area is an example of 3 stacking selections. The middle panel is a zoom on the gray area. Vertical dashed lines encompass the three selections of lines to be stacked. The vertical spikes are C$\alpha$ RRLs. The bottom panel shows the three stacks computed from the selections. Stacking allows for the detection of three velocity components{. Their expected positions are determined from previous studies \citep{Oonk14,Oonk17,Salas17} and} are marked by the vertical coloured lines: blue (--47~km~s$^{-1}$), green (--38~km~s$^{-1}$) and red (0~km~s$^{-1}$). The mean S/N for these three stacks is about $\sim$100 at $\sim$1~{k}m~s$^{-1}$ resolution.}
    \label{fig:post-proc}
\end{figure*}

\subsection{Stacking}
\label{subsub:stacking}
The goal of stacking is to increase further the S/N. However, as the shape (both peak and width values) of the lines depends on the frequency of the transition (see Sect. \ref{sub:line-theory}), the stacking procedure (see Fig.~\ref{fig:stacking}) tends to cancel out the specificities of each transition. This step is hence a balance between a sufficient S/N to ensure a detection, and as little stacking as possible to keep the line profiles intact.
The number of lines included inside the stacks depends on the source and on the frequency, as the rms of the data is higher on the edges of the spectrum. To maintain the same S/N throughout the whole spectrum, we thus need to stack together more lines at lower frequencies. 

We preselect a generic set of subsequent lines, whose size depend on the frequency. We then filter out heavily polluted lines (by RFI or a poor bandpass), to prevent propagation of artifacts in the final stack.
To automatically detect these heavily polluted lines, we applied the following procedure to the preselected lines for the stacking:
\begin{itemize}
    \item Each line of the preselection is convolved with a generic Voigt profile, whose characteristics (width and depth) depends on the source. 
    \item The convolved lines are interpolated to a common reference grid. This step ensures that all lines are aligned on the same velocity scale, facilitating accurate comparison and combination.
    \item Each convolved line is compared to a median reference line derived from the preselection. To assess the similarity between each convolved line and the reference, we use the $R^2$ coefficient of linear correlation, and the ratio between the convoluted line and the reference.
    \item Lines that show poor correlation ($R^2 < 0.7$ or ratio greater than 1.3) with the reference (based on the similarity assessment) are identified as outliers. These outliers are excluded from further analysis to prevent them from skewing the results.
    \item After removing outliers, a new median reference line is calculated from the remaining lines. This refined median line represents a cleaner and more accurate average of the spectral lines.
    \item The process is reiterated with the refined median reference line until no outliers are detected.
\end{itemize}
Finally, a visual assessment was applied on the remaining lines, in order to ensure the highest quality for the detections. Towards Cassiopeia A, the lines removed through visual assessment account for $\sim$30\% of the discarded lines. Towards Cyg A, the S/N was too low for non-stacked spectra, making it irrelevant to try and convolve them with a Voigt profile. We applied only the visual assessment for this source.

We eventually applied the stacking procedure. First each individual line is converted from a frequency axis to a velocity axis. Then, every line within a given stack is interpolated to a common velocity grid, defined by the coarsest grid. Finally, the isolated lines are weighted by the inverse square of their rms noise values. In order to account for the uncertainty on the calibration of NenuFAR, we determined a minimal value for the rms, which was obtained by capping the S/N at a value of 6. Lines presenting a better S/N were kept, but with a weight value equivalent to S/N of 6. The weighted lines are then summed to obtain an average spectral line profile. The process of stacking is illustrated for Cassiopeia A in Fig.~\ref{fig:post-proc}. The exact number of lines used in each stacks can be found in Table~\ref{tab:stacks}.

\section{Results}
\label{sec:results}
\begin{figure*} 
    \includegraphics[width=\textwidth]{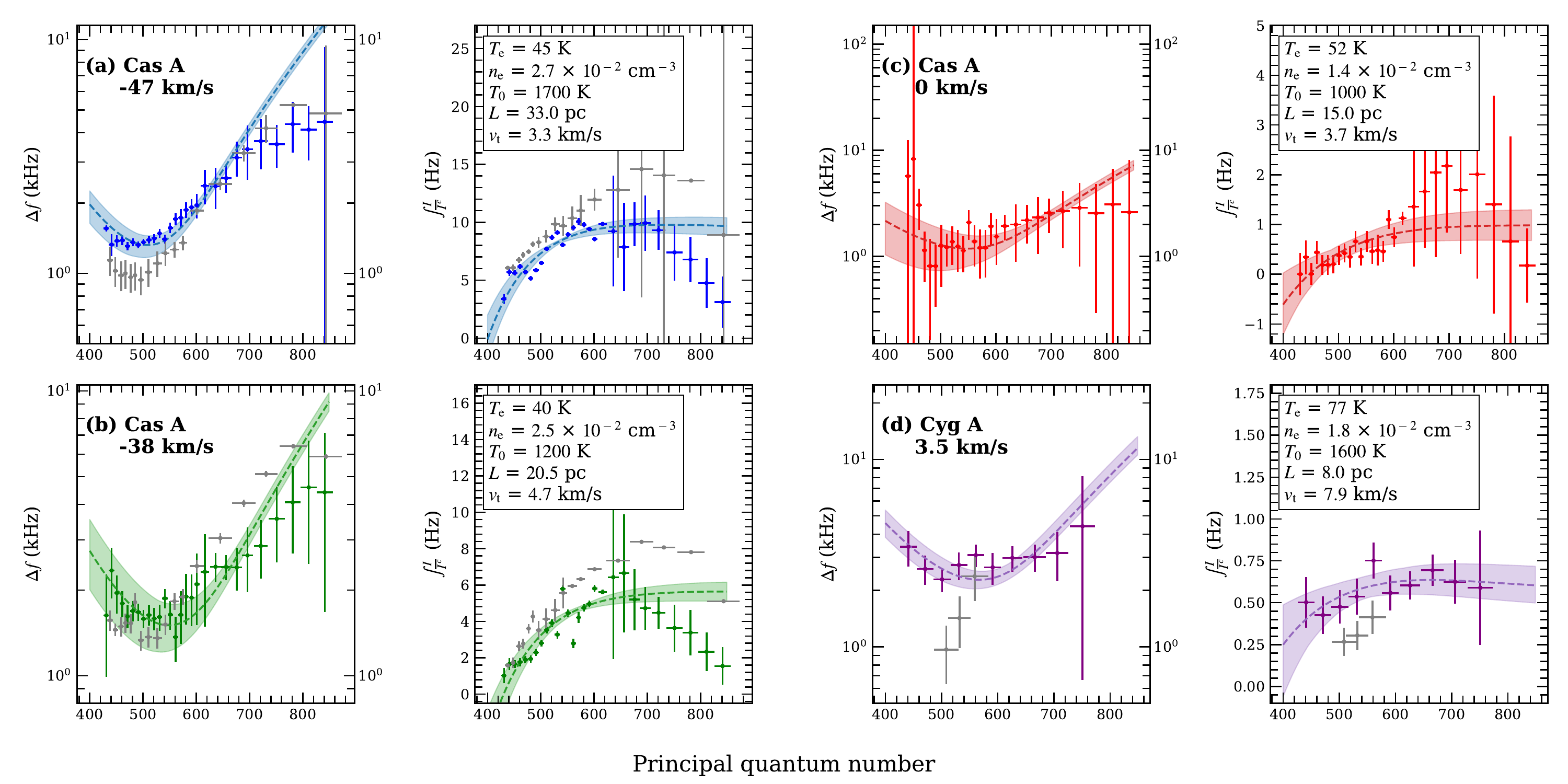}
    \caption{Fit results for the clouds detected towards Cas A and Cyg A. The coloured points correspond to NenuFAR data obtained through LT10. The gray points correspond to LOFAR Data from \citet{Oonk14,Salas17}. The coloured areas correspond to uncertainty range, determined as every parameter set that gives a distance within the lowest 30\%. The dashed and coloured lines are the models for the optimal parameters of the minimization. The boxed parameters are the optimal values of the minimization.}
    \label{fig:result-all-clouds}
\end{figure*}

From the stacked lines obtained from the procedure described in Sect. \ref{sec:redpipe}, we infer 5 physical parameters for the diffuse clouds in the line-of-sight: the electron temperature $T_\mathrm{e}$, the electron density $n_\mathrm{e}$, the temperature of the radiation field $T_0$, the typical size of the cloud $L$ and the mean turbulent velocity $\varv_\mathrm{t}$. The procedure used to extract these parameters from the stacked lines is described in Appendix~\ref{sec:fitlin}.
This section presents the results of these procedures for both sources: Cas A (Sect. \ref{sub:casao}) and Cyg A. (Sect. \ref{sub:cygao}). For each source, we subsequently present the characteristics of the detected lines (in Sects. \ref{subsub:casao-detections},\ref{subsub:cygao-detections}), the results of the line fitting (in Sects. \ref{subsub:casao-linefitting},\ref{subsub:cygao-linefitting}) and the evaluation of the physical parameters (in Sects. \ref{subsub:casao-physics},\ref{subsub:cygao-physics}). Our physical estimates with their uncertainties can be found in Table \ref{tab:result-of-gridsearch}, and the optimisation results are represented in Fig. \ref{fig:result-all-clouds}.

\begin{table*}[h!]
    \centering
    \caption{Best fit and uncertainties of the grid exploration.}
    \begin{tabular}{|c|c|c|c|c|c||c|c|c|}
    \hline
        Cloud   & $T_\mathrm{e}$ & $n_\mathrm{e}$ & $T_0$\tablefootmark{(a)} & $L$\tablefootmark{(a)}  & $\varv_\mathrm{t}$ & $p_{\rm th}/k$  &   $p_{\rm t}/k$    & EM$_\ion{C}{II}$ \\
             & [K]            & [$\cmcube$]    & [K]                      & [pc]                    & [km~s$^{-1}$]     & [$10^3$~K$\cmcube$] & [$10^5$~K$\cmcube$]   & [cm$^{-6}$ pc]   \\ \hline
                        
        Cas A -47     & $45\substack{+5\\-10}$   & $2.7\substack{+1.2\\-0.0}\times 10^{-2}$ & $1700\pm 100$  & $33.0\pm 0.5$ & $3.3\substack{+0.5 \\ -0.6}$  & 6.8 $-$ 13.9 & 0.6$-$6.5  & 0.024 $-$ 0.051 \\
        Cas A -38     & $40\substack{+4\\-3}$    & $2.5\substack{+0.0\\-0.0}\times 10^{-2}$ & $1200\pm 100$  & $20.5\pm 0.5$ & $4.7\substack{+1.4 \\ -1.3}$  & 6.6 $-$ 7.9  & 0.8$-$10.7 & 0.012 $-$ 0.013 \\
        Cas A  0      & $52\substack{+19\\-7}$   & $1.4\substack{+0.1\\-0.1}\times 10^{-2}$ & $1000\pm 100$  & $15.0\pm 0.5$ & $3.7\substack{+1.9 \\ -2.0}$  & 4.2 $-$ 7.6  & 0.1$-$7.6  & 0.002 $-$ 0.003 \\
        Cyg A 3.5    & $77\substack{+82\\-20}$   &{ $1.8\substack{+0.9\\-0.2}\times 10^{-2}$ }& $1600\pm 100$ &  $8.0\pm 0.5$ & $7.9\substack{+1.4 \\ -1.3}$   & 6.5 $-$ 30.7 & 2.0$-$27.0 & 0.002 $-$ 0.006 \\
        \hline
    \end{tabular}
    \tablefoot{The uncertainties on $T_{\rm e}$, $n_{\rm e}$ and $\varv_{\rm t}$ are computed by taking the sets of parameters giving that are within 30\% of the optimal $\chi^2$. \\
    \tablefoottext{a}{The bounds on $T_0$ and $L$ are fixed around the optimal value for each, as they are poorly constrained in the adopted modelling.} }
    
    \label{tab:result-of-gridsearch}
\end{table*}

\subsection{Cassiopeia A}
\label{sub:casao}
\subsubsection{Detections}
\label{subsub:casao-detections}
Cas A is the brightest radio source in the northern sky at 50 MHz \citep{deGasperin20}. It was observed by NenuFAR LT10 for 71.5 hours. Towards this source, we detected 398 individual C$\alpha$ lines between $n=426$ and $n=826$. Figure \ref{fig:CasA-all-detections-lines} shows the resulting stacked spectra. The three known ISM clouds in the foreground of Cas A at radial velocities of --47, --38 (with typical optical depths of a few $10^{-3}$) and 0~km~s$^{-1}$ (with typical optical depths of a few 10$^{-4}$, \citealt{Oonk17,Salas17}) are detected, at these opacity levels with NenuFAR. The --47 and --38~km~s$^{-1}$ components are variably blended throughout the observed band. This is due to a combination of two effects: the broadening of the lines at low frequencies (see Sect. \ref{sub:line-theory}), and the dependency on the frequency of the conversion between frequency and velocity channels.  
The blending of the components, which occurs at frequencies lower than 24.34~MHz, makes it difficult to disentangle the absorption from these two components over a fraction of the observed band. We took these effects into account when performing the stacking described in Sect. \ref{subsub:stacking}. For this, we stacked together groups of 10 to 30 transitions depending on the frequency (see Table \ref{tab:stacks}).
To assess the robustness of our methodology, we compared our detections with the LOFAR LBA ones \citep{Oonk17,Salas17}. Our study significantly increases the S/N in comparison with those, as can be seen in Figs.~\ref{fig:nenufar-vs-lofar;cas,oonk} and \ref{fig:nenufar-vs-lofar;cas,salas}. This allows the detection of the 0~km~s$^{-1}$ component on a broader range of frequencies, and with less stacking needed. The S/N is increased by a factor of about 4 at the same velocity resolution, and by a factor of about 2 with the same integration time. 
We note a discrepancy between LOFAR and NenuFAR in the integrated line-to-continuum intensity of the lines (see Figs.~\ref{fig:nenufar-vs-lofar;cas,oonk} and \ref{fig:nenufar-vs-lofar;cas,salas}). This difference is above the noise, and varies from a few percent at high frequencies to a few tens of percent at low frequencies. We discuss it in Sect.~\ref{sub:beam-effects}. 

\begin{figure*}[ht]
\centering
    \begin{subfigure}[b]{0.6\textwidth}
        \centering
        \includegraphics[height=10cm]{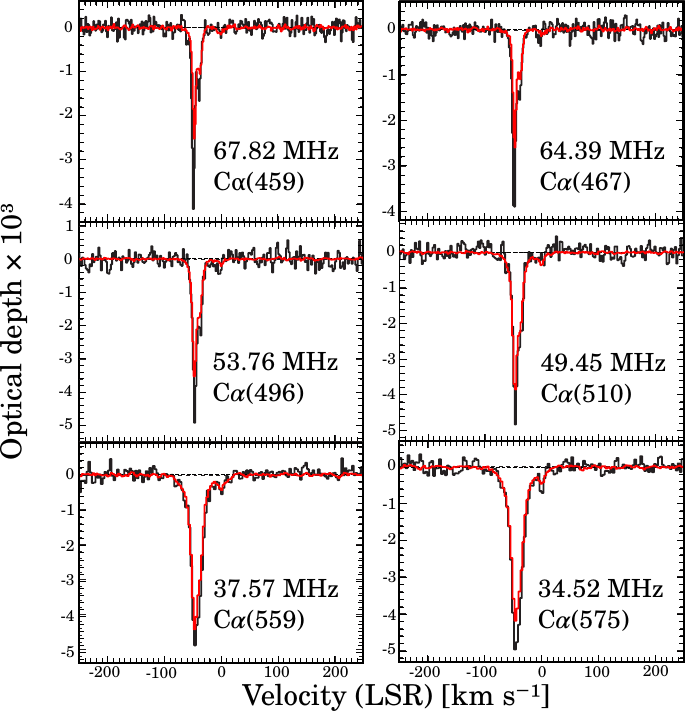}
        \caption{\citet{Oonk14}: LOFAR LBA 35 $-$ 80 MHz}
        \label{fig:nenufar-vs-lofar;cas,oonk}
    \end{subfigure}
    \begin{subfigure}[b]{0.35\textwidth}
        \centering
        \includegraphics[height=10cm]{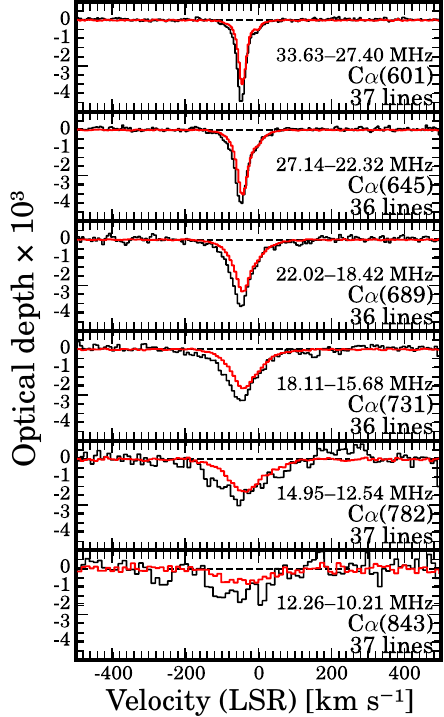}
        \caption{\citet{Salas17}: LOFAR LBA 10-35 MHz}
        \label{fig:nenufar-vs-lofar;cas,salas}
    \end{subfigure}
    ~
    \begin{subfigure}{0.95\textwidth}
        \includegraphics[width=\textwidth]{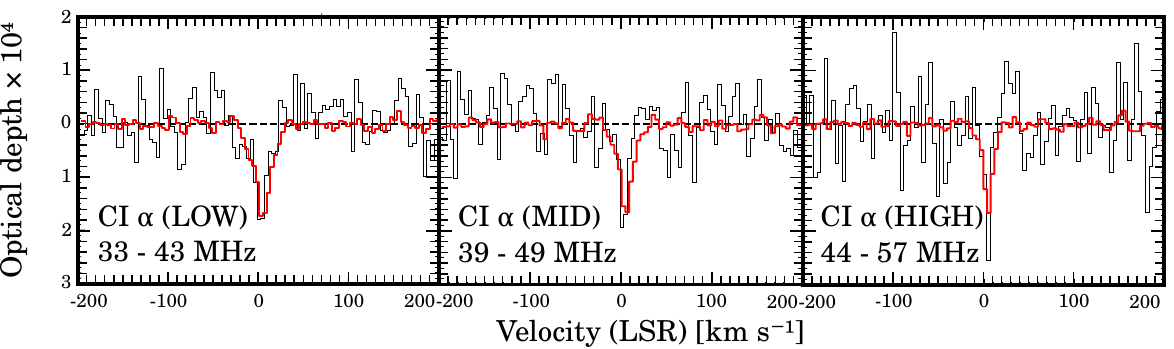}
        \caption{{\citet{Oonk14}: LOFAR LBA 33-57 MHz}}
        \label{fig:nenufar-vs-lofar;cyg}
    \end{subfigure}
    \caption{Overplot of the lines detected towards Cas A (\ref{fig:nenufar-vs-lofar;cas,oonk}, \ref{fig:nenufar-vs-lofar;cas,salas}) and Cyg A (\ref{fig:nenufar-vs-lofar;cyg}) for LOFAR LBA ({black}) and NenuFAR (red). The panels \ref{fig:nenufar-vs-lofar;cas,oonk} compare the detections of LOFAR LBA in the frequency range $\sim$35 to $\sim$68~MHz, from \citet{Oonk17}. The panels \ref{fig:nenufar-vs-lofar;cas,salas} compare the detection of LOFAR LBA from $\sim$10 to $\sim$35~MHz, from \citet{Salas17}. The panels \ref{fig:nenufar-vs-lofar;cyg} compare the detection of LOFAR LBA from $\sim$35 to $\sim$57~MHz, from \citet{Oonk14}. The top panel (LOW) presents the detections of LOFAR LBA in the frequency range $\sim$33 to $\sim$43~MHz, the middle panel (MID) presents the detection from $\sim$39 to $\sim$49~MHz and the bottom panel (HIGH) is the detections from $\sim$44 to $\sim$57~MHz.
    The detections from NenuFAR were rebinned to the same velocity resolution as LOFAR, and the same transitions were included in the stacks.}
    \label{fig:nenufar-vs-lofar;cas;cyg}
\end{figure*}

\subsubsection{Line fitting and cloud identification}
\label{subsub:casao-linefitting}
For Cassiopeia A, the primary approach is to fit simultaneously three Voigt profiles to account for the three velocity components. Two issues arise regarding the fitting.
First, the depth of the CRRLs varies with frequency as per the theory, and this variation is different for each different velocity clouds, as they do not necessarily share the same physical parameters. This means that some velocity component might dominate the others in some parts of the spectrum.
Moreover, as the size of the velocity channels widens as the frequency decreases, the velocity components blend below a certain frequency (around 40~MHz) until they are virtually indistinguishable (around 30~MHz).
To manage these specificities, we adjusted the fitting prior and search bounds according to the principal quantum number, in order to provide a made-to-measure guide for the fitting algorithm:
\begin{itemize}
    \item For $n$ below 560 (\textit{i.e.} $\nu>37.37$~MHz), the centre of the Voigt profiles of the -47~km~s$^{-1}$ component was confined between $-50$ and $-44$~km~s$^{-1}$. The centre of the two other components were defined by their distance from the -47~km~s$^{-1}$ component: the -38~km~s$^{-1}$ and the 0~km~s$^{-1}$ must be respectively at $(+9\pm2)$~km~s$^{-1}$ and $(+47 \pm 2)$~km~s$^{-1}$ of the main velocity component. 
    \item For $n$ above 560, we keep the same bound on the centre of the main Voigt profile, but the distance with the other components can vary by only 1~km~s$^{-1}$. Moreover, we force the line area of the -38~km~s$^{-1}$ component to be less or equal to a third of the line area of the -47~km~s$^{-1}$. The line area of the 0~km~s$^{-1}$ must be less than or equal to the -38~km~s$^{-1}$ component. This strategy is tailored to mitigate as much as possible the effect of the blending of the -47 and -38~km~s$^{-1}$. We also force a dynamic bound on the width of the lines: the Gaussian width must be almost constant (as the Gaussian width is supposed to be independent from the frequency, see Sect.~\ref{sub:line-theory}), and the Lorentzian width must be globally increasing.
\end{itemize}
Due to the blending, the uncertainties above $n\sim 600$ are dominated by the inability for the \texttt{curve\_fit} to differentiate properly the velocity components. To assess the uncertainty on the linewidth and the line integrated intensity, we applied a statistical method. 
A stacked line is composed of a set of consecutive transitions. For each stack, we selected 300 random subsets of transitions, of varying lengths, that we stacked and then fitted with the same methodology as described above. We extracted from these fits a value of linewidth and integrated intensity for the 3 components. We then evaluated the linewidths and integrated intensities and their uncertainties for each stack as the mean and  3~rms of the values from every subsets.
We applied this methodology for each stacked line whose average quantum number is above 600. The result{s} of the line fitting are available in Fig.~\ref{fig:CasA-all-detections-lines} and in Table~\ref{tab:result-casa}. 

\begin{figure}[h!]
    \centering
    \includegraphics[width=0.9\linewidth]{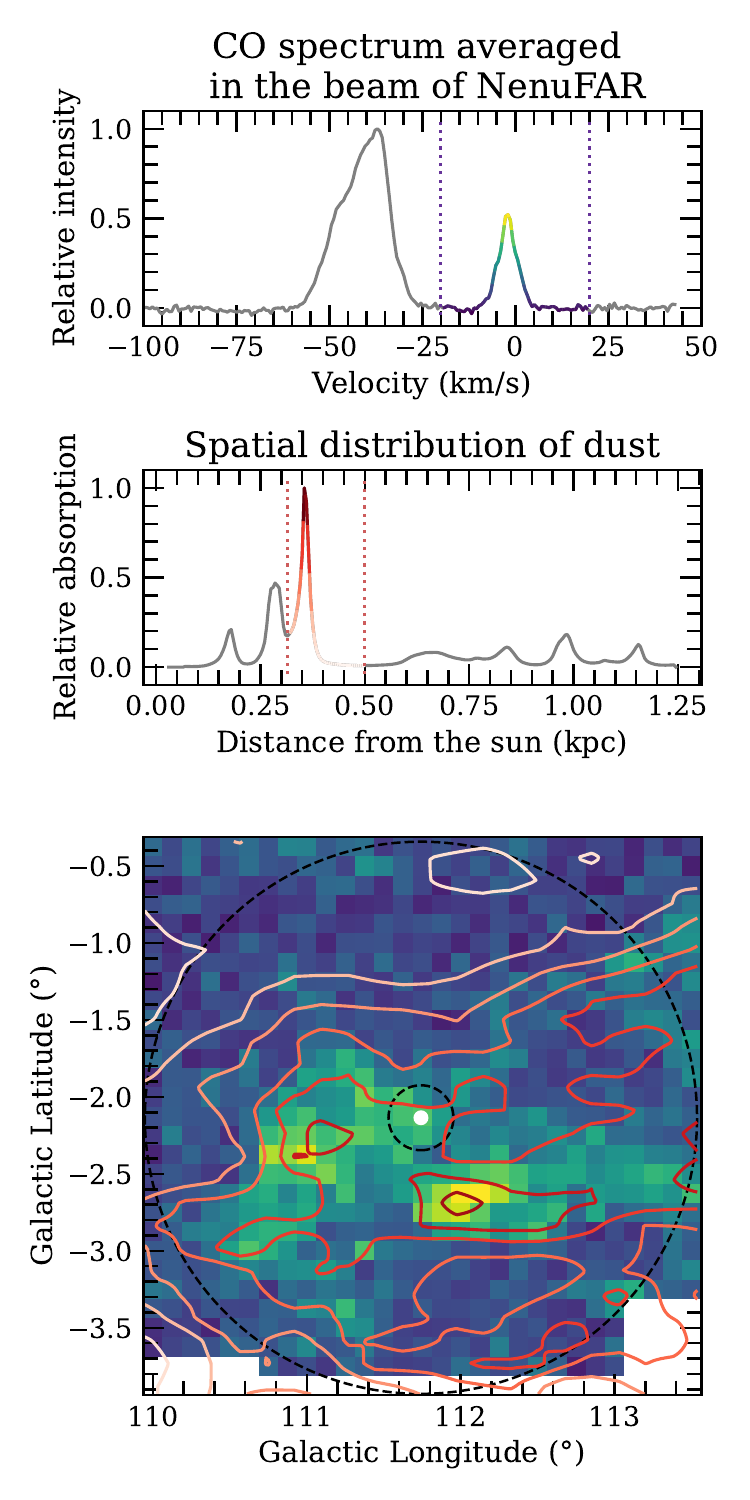}
    \caption{Cloud identification for the 0~km~s$^{-1}$ component of Cas A.
    Top panel: CO spectrum averaged over the largest beam of NenuFAR. The gray line represents the whole spectrum. The coloured part is the sliced used to draw the intensity map in the bottom panel \citep[see][]{Dame01}. Middle panel: dust distribution along the sightline of Cas A \citep[see][]{Edenhofer24}. The gray line represents the sightline up to a distance of 1.25 kpc. The coloured part is the slice used to plot the contours in the bottom panel.
    Bottom panel: overplot of the intensity of CO and dust absorption in the plane of sky. The colour map is the 0$^{\rm th}$ moment of the CO cube limited to velocities between -20 and 20~km~s$^{-1}$. The contours are the absorption of the dust located between 0.3 and 0.5~kpc from the Sun. The white filled circle represents the position and size of Cas A. The black dashed circles are the largest and smallest beams of NenuFAR (respectively 3.59$^\circ$ and 24.4').}
    \label{fig:cloud-identification-0}
\end{figure}

\begin{figure}[h!]
    \centering
    \includegraphics[width=0.9\linewidth]{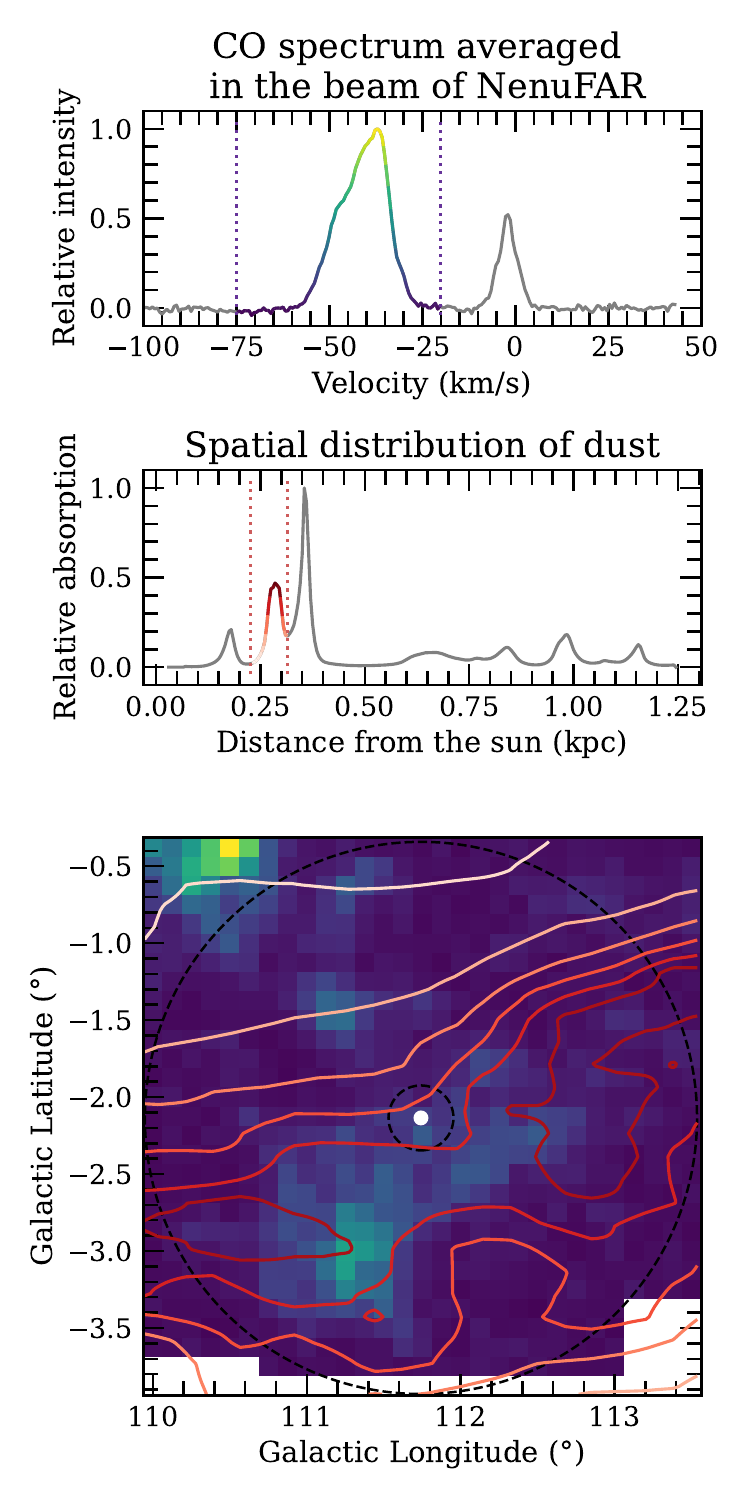}
    \caption{Same as Fig. \ref{fig:cloud-identification-0} but for the -47 and -38~km~s$^{-1}$ component of Cas A.}
    \label{fig:cloud-identification-4738}
\end{figure}

{Following the approach described in Sect.~\ref{subsub:cloud-id}, we performed a cross-analysis of the CRRLs with dust clouds and CO in the corresponding line-of-sight.} The dust distribution in the sightline of Cas A up to 1.25~kpc presents 3 main absorption peaks, at about 175, 280 and 350~pc from the Sun (middle panels of Figs. \ref{fig:cloud-identification-0},\ref{fig:cloud-identification-4738}). The CO spectrum averaged in a beam of 3.59$^\circ$ (corresponding to the beam of NenuFAR at 10~MHz) presents two components: a wider and stronger one between $\sim-60$ and $\sim-20$~km~s$^{-1}$ with a central velocity around -40~km~s$^{-1}$, and another one wide of about 15~km~s$^{-1}$ and with a central velocity of $-3$~km~s$^{-1}$ (top panels of Figs. \ref{fig:cloud-identification-0},\ref{fig:cloud-identification-4738}). The largest velocity component seems to harbour two distinct components, peaking at $-40$ and $-50$~km~s$^{-1}$. We thus relate these components respectively to the $-38$ and $-47$~km~s$^{-1}$ clouds detected with CRRLs. 
We identify a spatial correlation in the plane-of-sky between the dust cloud located at a distance of 282~pc from the Sun, and the $-47$ and $-38$~km~s$^{-1}$ velocity components, as shown in Fig. \ref{fig:cloud-identification-4738}. We measure a FWHM of 24~pc along the line-of-sight for this dust cloud.  We were unable to identify two separate spatial structure corresponding to the two velocity components, and assumed the same prior $L$ for both.
Using the same methodology, we identify a correlation between the dust cloud located at 350~pc from the Sun and the 0~km~s$^{-1}$ velocity component of Cas A, as shown in Fig. \ref{fig:cloud-identification-0}. We measure a FWHM of 8~pc for this component. 

\subsubsection{Determination of the physical parameters}
\label{subsub:casao-physics}
For the -47~km~s$^{-1}$ velocity component, we found a solution with optimal values estimated at $n_\mathrm{e}=2.7\times10^{-2}-\SI{3.9e-2}{\per\centi\meter\cubed}${, with $n_\mathrm{e,opt}= 2.7 \cmcube$}, $T_\mathrm{e} = 35-50$~K{, with $T_\mathrm{e,opt}= 45$~K} and $\varv_t=2.7-3.8$~km~s$^{-1}${, with $\varv_\mathrm{t,opt}= 3.3$~km~s$^{-1}$}. This solution is found for $T_0 = (1700\pm100)$~K and $L=32.5-33.5$~pc.
From these values, we infer EM$_\ion{C}{II}=0.024-0.051$~cm$^{-6}$~pc, $p_{\rm th}/k=6.8-\SI{13.9e3}{\kelvin\per\centi\meter\cubed}$ and $p_{\rm t}/k=0.6-\SI{6.5e5}{\kelvin\per\centi\meter\cubed}$. These values are consistent with the prediction of \citet{Wolfire03} for the neutral ISM of the Milky Way, of a thermal pressure $p_{\rm th}\sim \SI{10e3}{\kelvin\per\centi\meter\cubed}$ that is dominated by the turbulent pressure. We also infer the total Doppler broadening to be between 4.5 and 6.3~km~s$^{-1}$.
Overall, the measured parameters point towards a diffuse partially ionised phase of the ISM. The result of the grid search is presented in Fig.~\ref{fig:casa-47-histo}. An overplot of the data points and the best fit is in panel~(a) of Fig.~\ref{fig:result-all-clouds}. 

The values inferred by \citet{Salas17} for this cloud are $n_\mathrm{e}=2\times10^{-2}-\SI{3.5e-2}{\per\centi\meter\cubed}$, $T_\mathrm{e} = 60-98$~K, $T_0 = 1500 - 1650$~K and $p_{\rm th}/k=11 - \SI{21e3}{\kelvin\per\centi\meter\cubed}$. 

\citet{Oonk17} inferred  $p_{\rm th}/k=19 - \SI{29e3}{\kelvin\per\centi\meter\cubed}$, $p_{\rm t}/k=1.8-\SI{2.0e5}{\kelvin\per\centi\meter\cubed}$, a total Doppler broadening of 3.4~km~s$^{-1}$, EM$_\ion{C}{II} = 0.042 - 0.070$ \si{\per\centi\meter^{-6}~pc}, $n_\mathrm{e}=3.5\times10^{-2}-\SI{4.5e-2}{\per\centi\meter\cubed}$, $T_\mathrm{e} = 80-90$~K, $T_0 = 1268 - 1484$~K and $L=34.1 - 36.5$~pc. 
Our estimates are in general agreement with LOFAR results. The difference in Doppler broadening is understandable due to the dependency of $\varv_\mathrm{t}$ on the beam size (see Sect.\ref{sub:beam-effects}). The slight differences between LOFAR and NenuFAR estimates are sufficiently small as to not impact the interpretation of the component as a diffuse partially ionised medium.

Contrary to \citet{Salas17}, who assumed the same physical properties for the two components, NenuFAR's higher spectral resolution and S/N allowed us to distinguish differences between the $-$47 and $-$38 ~km~s$^{-1}$ clouds, 
For the latter, we found a solution with optimal values estimated at $n_\mathrm{e}=\SI{2.5e-2}{\per\centi\meter\cubed}$, $T_\mathrm{e} = 37-44$~K{, with $T_\mathrm{e,opt}= 40$~K} and $\varv_t=3.4-6.1$~km~s$^{-1}${, with $\varv_\mathrm{t,opt}= 4.7$~km~s$^{-1}$}. This solution is found for $T_0 = (1200\pm100)$~K and $L=20-21$~pc.
From these values, we infer EM$_\ion{C}{II}=0.012-0.013$~cm$^{-6}$~pc, $p_{\rm th}/k=6.6-\SI{7.9e3}{\kelvin\per\centi\meter\cubed}$, $p_{\rm t}/k=0.8-\SI{10.7e5}{\kelvin\per\centi\meter\cubed}$ and a Doppler broadening between 5.7 and 10.8~km~s$^{-1}$. The thermal pressure inferred here is slightly below the expected value of $\sim\SI{10e3}{\kelvin\per\centi\meter\cubed}$. The turbulent pressure dominates over the thermal pressure, as expected. Again, our solution points to a diffuse and partially ionised medium. 
\citet{Oonk17} found $p_{\rm th}/k=19 - \SI{29e3}{\kelvin\per\centi\meter\cubed}$, $p_{\rm t}/k=6.6-\SI{8.6e5}{\kelvin\per\centi\meter\cubed}$, a total Doppler broadening of 6.8~km~s$^{-1}$, EM$_\ion{C}{II} = 0.022 - 0.038$ \si{\per\centi\meter^{-6}~pc}, $n_\mathrm{e}=3.5\times10^{-2}-\SI{4.5e-2}{\per\centi\meter\cubed}$, $T_\mathrm{e} = 75-95$~K, $T_0 = 1379 - 1635$~K and $L=17.0 - 20.2$~pc. \par
Although there is a slight deviation between LOFAR's and NenuFAR's estimates in the analysis of both clouds, these differences are understandable as the data points from NenuFAR deviate from the ones from LOFAR (see Fig. (b) of \ref{fig:result-all-clouds}). The differences in estimate are sufficiently small as to not impact the interpretation of the component as a diffuse partially ionised medium. We discuss this issue further in Sect. \ref{sub:beam-effects}. \par
An overplot of the data points and the best fit is in Fig.~\ref{fig:casa-38-histo}. The optimal modelling fits well the linewidth for quantum numbers below $n\sim 600$ and deviates afterwards. The uncertainty area generally encompasses the integrated area data points below $n\sim 600$. Above $n\sim 600$, the measure of the integrated line-to-continuum intensity collapses. This is due to the blending of the components, and the fact that the algorithm only sees one Voigt profile instead of two. In addition, at these quantum numbers, the Lorentzian wings are wider and blend with the continuum, resulting in the algorithm potentially underestimating their contribution to the intensity. Thus, the decrease in line intensity is not physical, and cannot be reproduced by our modelling, so it should not impact the optimisation process. Figure (b) of \ref{fig:result-all-clouds}) shows indeed that the optimisation fitted mostly the left part of the curves.

For the 0~km~s$^{-1}$ velocity component, we found a solution with optimal values estimated at $n_\mathrm{e}=1.3\times10^{-2}-\SI{1.5e-2}{\per\centi\meter\cubed}${, with $n_\mathrm{e,opt}= 1.4~\cmcube$}, $T_\mathrm{e} = 45-71$~K{, with $T_\mathrm{e,opt}= 52$~K}, and $\varv_t=1.7-5.6$~km~s$^{-1}${, with $\varv_\mathrm{t,opt}= 3.7$~km~s$^{-1}$}. This solution is found for $T_0 = (1000\pm100)$~K, $L=14.5-15.5$~pc. 
From these values, we infer EM$_\ion{C}{II}=0.002-0.003$~cm$^{-6}$~pc, $p_{\rm th}/k=4.2-\SI{7.6e3}{\kelvin\per\centi\meter\cubed}$, $p_{\rm t}/k=0.1-\SI{5.4e5}{\kelvin\per\centi\meter\cubed}$ and a Doppler width between 2.9 and 9.3~km~s$^{-1}$. The thermal pressure inferred here is slightly below the expected value of $\sim\SI{10e3}{\kelvin\per\centi\meter\cubed}$. The turbulent pressure is higher, but the extremal values of the pressures are separated from each other by less than an order of magnitude. This suggests that this cloud is at the limits of a typical behaviour of a CNM cloud.

Although NenuFAR is the first instrument allowing to place physical constraints on the 0~km~s$^{-1}$ component, \citet{Oonk17} was able to measure a Doppler width of $w_\mathcal{D}=2.5$~km~s$^{-1}$, slightly lower than our measure. The results of the grid search can be found in Fig.~\ref{fig:casa-0-histo}, and an overplot of the data points and uncertainty ranges in Fig. (c) of \ref{fig:result-all-clouds}.

\subsection{Cygnus A}
\label{sub:cygao}
\subsubsection{Detections}
\label{subsub:cygao-detections}
{Cyg A was} observed by NenuFAR LT10 for 157.5 hours. There is one single known ISM cloud in the foreground of Cyg A at radial velocities of 3.5~km~s$^{-1}$ \citep{Oonk14}, with typical optical depths of a few $10^{-4}$. This is why we chose to stack together groups of 30 to 50 transitions depending on the frequency, ending up in eleven usable detections. The results are shown in Fig.~\ref{fig:CygA-detections-lines}, and the characteristics of the detected lines are listed in Table \ref{tab:stacks}. We compare our results with the LOFAR LBA ones from \citep{Oonk14}. As for Cas A, our study improves the S/N (see Fig.~\ref{fig:nenufar-vs-lofar;cyg}). When stacking the exact same transitions as \citet{Oonk14}, and rebinning our data to the LOFAR LBA frequency resolutions, we obtained a higher S/N than the LOFAR LBA one by a factor 10, and by a factor of 3 for the same integration time. In terms of optical depth, we note that the discrepancy between LOFAR LBA and NenuFAR is much less important towards Cyg A than towards Cas A (see Fig.~\ref{fig:nenufar-vs-lofar;cyg}), for reasons that we discuss in Sect.~\ref{sec:discuss}).
\subsubsection{Line fitting and cloud identification}
\label{subsub:cygao-linefitting}
\begin{figure}[h!]
    \centering
    \includegraphics[width=\linewidth]{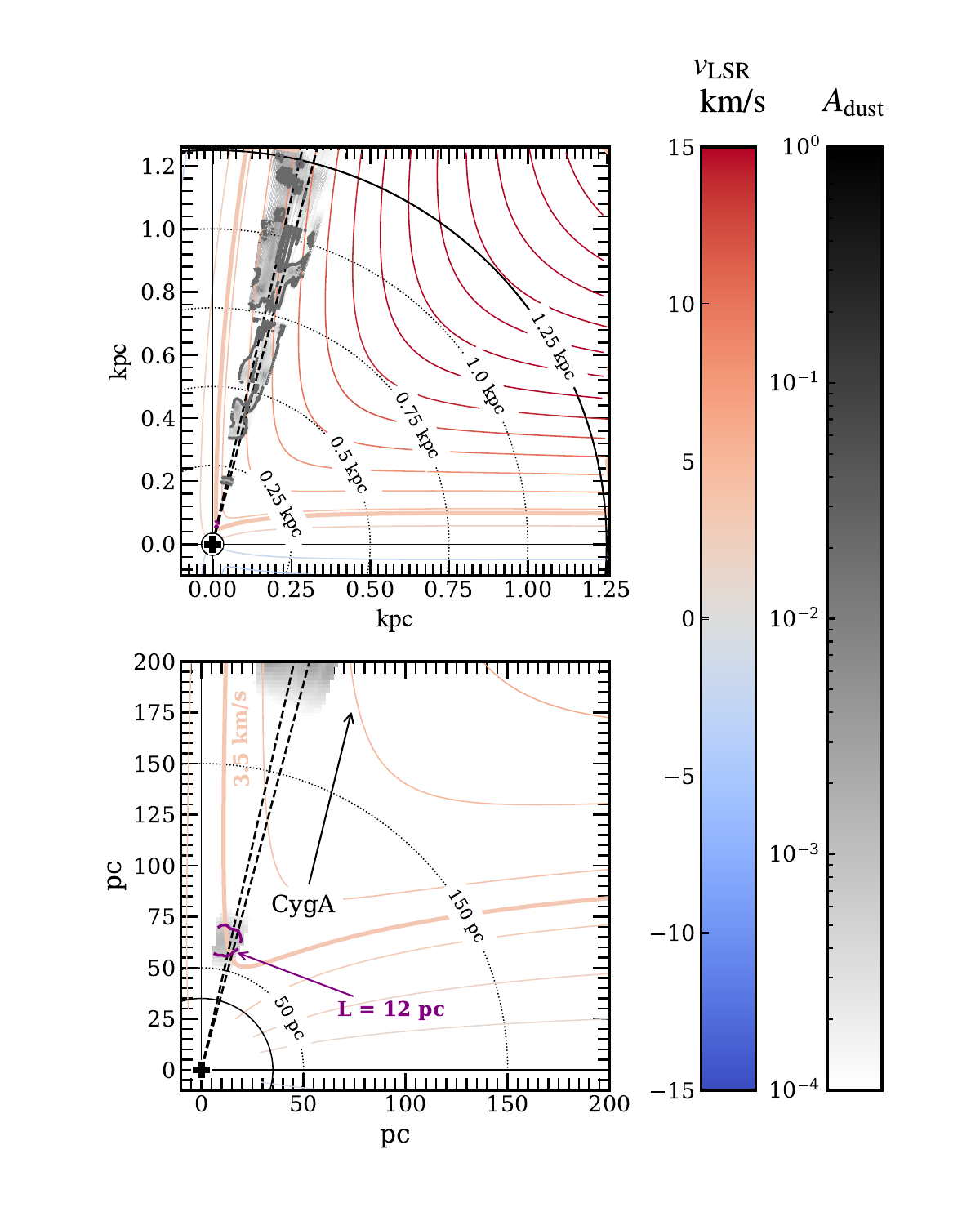}
    \caption{Cloud identification for the 3.5~km~s$^{-1}$ component towards Cyg A. 
    Both panels are an overplot of the dust distribution in gray \citep{Edenhofer24} and theoretical LSR velocities in blue and red contours \citep{Reid19}. The largest beam of NenuFAR is marked in black dotted lines. The Sun is represented by the black cross.
    Top panel: overview of the sightline up to 1.25~kpc from the Sun. The gray contours correspond to the most prominent clouds in the beam of NenuFAR.
    Bottom panel: Zoom on the close vicinity, up to 200~pc from the Sun. The purple cloud crosses the beam towards Cyg A at the expected velocity of 3.5~km~s$^{-1}$.}
    \label{fig:cloud-identification-cyg}
\end{figure}

For Cygnus A, we rebinned every spectrum by a factor of 4, to improve the S/N.
The fit was achieved blindly, without prior knowledge or bounds.
The result{s} of the line fitting are available in Fig.~\ref{fig:CygA-detections-lines}.
The cloud identification method (see Sect.~\ref{subsub:cloud-id}) revealed no CO cloud towards Cyg A, but the dust map analysis revealed a cloud crossing the line-of-sight at the LSR velocity of 3.5~km~s$^{-1}$. This cloud has an extinction FWHM of 12~pc (see Fig.~\ref{fig:cloud-identification-cyg}), and is located at 70~pc from the Sun. We used this value as a prior on $L$ for the grid exploration. 

\subsubsection{Determination of the physical parameters}
\label{subsub:cygao-physics}
{We found a solution with optimal values estimated at $n_\mathrm{e}=1.6\times10^{-2}-\SI{2.7e-2}{\per\centi\meter\cubed}$, with $n_\mathrm{e,opt}= 1.8~\cmcube$, $T_\mathrm{e} = 57-159$~K{, with $T_\mathrm{e,opt}= 77$~K}, and $\varv_t=6.6-9.3$~km~s$^{-1}${, with $\varv_\mathrm{t,opt}= 7.9$~km~s$^{-1}$}. This solution is found for $T_0 = (1600\pm100)$~K, $L=7.5-8.5$~pc.
From these values, we infer EM$_\ion{C}{II}=0.002-0.006$~cm$^{-6}$~pc, $p_{\rm th}/k=6.5-\SI{30.7e3}{\kelvin\per\centi\meter\cubed}$ and $p_{\rm t}/k=2.0-\SI{27.0e5}{\kelvin\per\centi\meter\cubed}$.} These values are consistent with the prediction of \citet{Wolfire03} for the neutral ISM of the Milky Way, of a thermal pressure $p_{\rm th}\sim \SI{10e3}{\kelvin\per\centi\meter\cubed}$ that is dominated by the turbulent pressure.
\citet{Oonk14} conducted an analysis of CRRLs towards Cyg A using LOFAR LBA, and found optimal values of $T_\mathrm{e} \sim 110$~K with uncertainty ranges from 50 to 500~K, which are compatible with our results. The electron density from LOFAR is found at $n_\mathrm{e}\sim\SI{6e-2}{\per\centi\meter\cubed}$ with uncertainties from 0.005 to 0.07~\si{\per\centi\meter\cubed}, encompassing our results. They estimated EM$_\ion{C}{II}\sim 0.001$~cm$^{-6}$~pc, giving the same order of magnitude as our study. They also found rough estimates of $T_0 \sim 2700$~K, a Doppler width of $\sim$10~km~s$^{-1}$ and $p_{\rm th}/k\sim\SI{40e3}{\kelvin\per\centi\meter\cubed}$. Given the range of the quantum numbers ($n\gtrsim 500$), they assumed the line broadening to be only in Lorentzian regime, therefore the turbulent velocity was estimated using observations at higher frequencies. However, the data points acquired by NenuFAR indicate that the shift from Doppler dominated to Lorentzian regime happens around $n=580$, indicating that LOFAR's data points were not completely in the Lorentzian dominated region.

\section{Discussion}
\label{sec:discuss}

There are several limitations to our study. We discuss them in this section. In Sect.~\ref{sub:beam-effects}, we focus on the consequences of the beam size differences between LOFAR and NenuFAR. In Sect~\ref{sub:multimodality}, we describe the multimodality and intrinsic degeneracies of the 5-parameter model we used to describe CRRLs. 

\subsection{Beam effects}
\label{sub:beam-effects}
\begin{figure}
        \includegraphics[width=\linewidth]{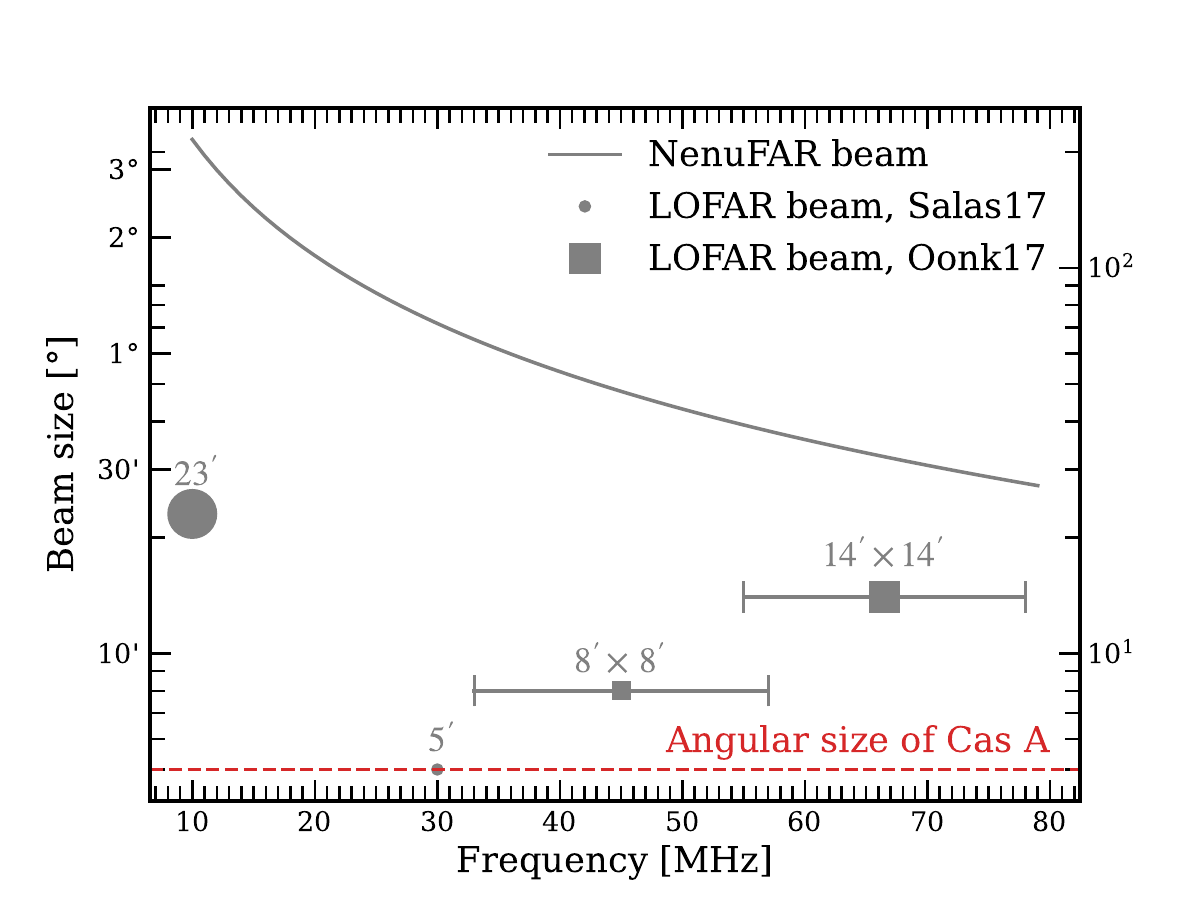}
        \caption{Frequency dependence of beam sizes for NenuFAR and LOFAR, shown alongside the angular size of Cas A. The LOFAR beam is represented for the two CRRLs experiment: \citet{Salas17} (circles) and \citet{Oonk17} (squares). \citet{Oonk17} cropped the images from LOFAR in squares whose sizes are fixed in two frequency windows. \citet{Salas17} cropped the images from LOFAR in circles centred on Cas A, with continuous values for its diameter. The NenuFAR beam size function is shown with the solid line. The red dotted line represents the angular extension of Cas A at radio frequencies \citep{Green14}. The source is spatially resolved for neither setups, at any frequency.}
        \label{subfig:beam-of-instruments}
\end{figure}

\begin{figure}
        \includegraphics[width=\linewidth]{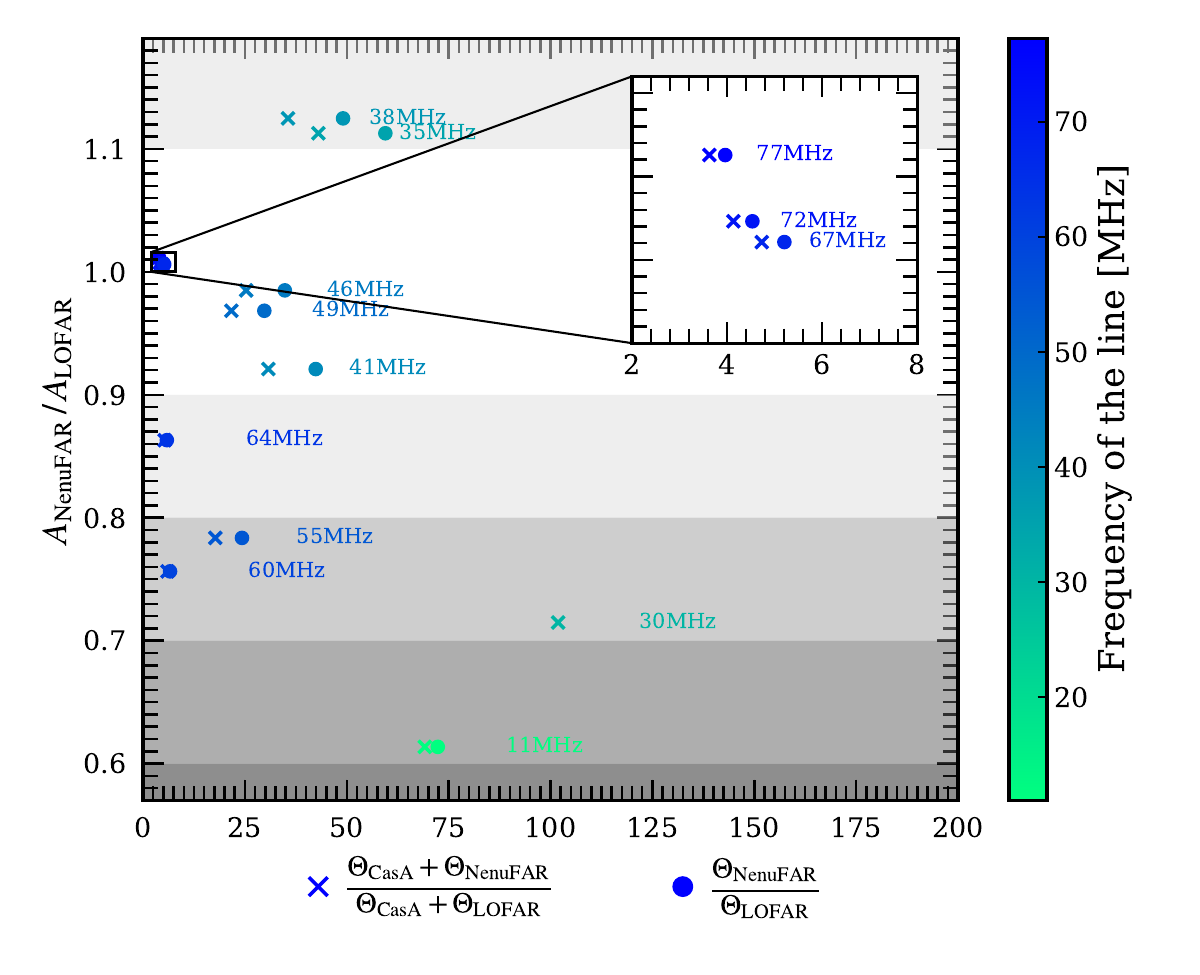}
        \caption{Ratio of line intensities towards Cas A between observations with NenuFAR and with LOFAR as a function of beam ratio. The crosses represent the filling factor ratios for Cas A and the dots represent the ratio of angular resolutions for the two experiments. If at a given frequency a dot and a cross are close to each other, then the filling factors are similar for both instruments, meaning a less prominent dilution effect between both experiments. The colour scale indicates the central frequency of the lines. The ratio of the line intensities follows the tendency of being farther from 1 at lower frequencies. The differences in filling factor follow the same tendency.}
        \label{subfig:beam-intensity-ratios}
\end{figure}

The discrepancy between the LOFAR and NenuFAR analyses (see Fig. \ref{fig:nenufar-vs-lofar;cas;cyg}) may be due to beam effects. In \citet{Salas17}, \citet{Oonk17} and \citet{Oonk14}, the observations were performed in imaging mode with a field of view varying from $5'$ to $23'$, depending on the frequency. The current beam of NenuFAR varies between $25.2'$ at \SI{85}{\mega\hertz} and $3.59^\circ$ at \SI{10}{\mega\hertz}. Figs.~\ref{subfig:beam-of-instruments} and \ref{subfig:beam-intensity-ratios} shows a quantitative analysis of beam (top panel) and filling factor (bottom panel) differences between NenuFAR and the different setups used with LOFAR. We identified three possible effects from these differences.

{First, the} observed mean turbulent velocity $\varv_\mathrm{t}$ increases with the size of the beam, as described by \citet{Gordon02}. This is because a larger beam encompasses a wider diversity of spatial structures hence of inner motions, leading on average to a higher turbulent velocity.
A greater mean turbulent velocity increases the Doppler broadening, and therefore raises the left part of the curve (corresponding to low $n$ values and high frequencies) of $\Delta \nu$ as a function of $n$ that corresponds to the Doppler-dominated frequencies. This is what we observed towards Cas A and Cyg A (see Fig. \ref{fig:result-all-clouds}). 

{Second, we suspect some dilution effect due to the differences in beam sizes. In order to observe CRRLs in absorption, the diffuse ISM needs to be in the foreground of a radio source. As the radio source is resolved in neither cases and for neither instruments (see Fig.~\ref{subfig:beam-of-instruments}), we would expect the lines to have a similar depth for both instrument. This is the case towards Cyg A (see Fig.~\ref{fig:nenufar-vs-lofar;cyg}), but not towards Cas A, especially at low frequencies (see Figs.~\ref{fig:nenufar-vs-lofar;cas,oonk} and Figs.~\ref{fig:nenufar-vs-lofar;cas,salas}). The difference between LOFAR and NenuFAR towards Cas A suggests that the background against which we see the CRRLs is actually resolved, and not similarly encompassed by the beams of the instruments. As this dilution effect is more prominent towards Cas A than Cyg A, the fact that Cyg A is located farther from the Galactic plane than Cas A, points that the Galactic Plane Background is a strong enough radio source for CRRLs detections. However, we suspect that the exact change in line strength between the instruments depends on the foreground and background gases configurations.}

Third, NenuFAR also presents grated lobes due to the hexagonal grid of the MAs (see Sect. \ref{sub:tnt}, Fig. \ref{fig:nenufar-layout}). The grating effect is more prominent at higher frequencies, with no effect at all visible below 27.3~MHz\footnote{see \href{https://nenufar.obs-nancay.fr/en/astronomer/}{https://nenufar.obs-nancay.fr/en/astronomer/}}. The grating lobes could affect the lines due to contamination by other bright sources that are not in the direction of the main target. As a result, new velocity component{s} could become visible to NenuFAR by entering the grating lobes throughout the night, which can induce line blending if the new velocities are too close to the ones we observe. However, Fig. \ref{subfig:beam-intensity-ratios} shows that the filling factors of NenuFAR and LOFAR (which does not present grating lobes) are closer at high frequency. We thus consider that the grating lobes effect is negligible in our case.

\subsection{Multimodality}
\label{sub:multimodality}
\begin{figure}[h!]
    \centering
    \includegraphics[width=\linewidth]{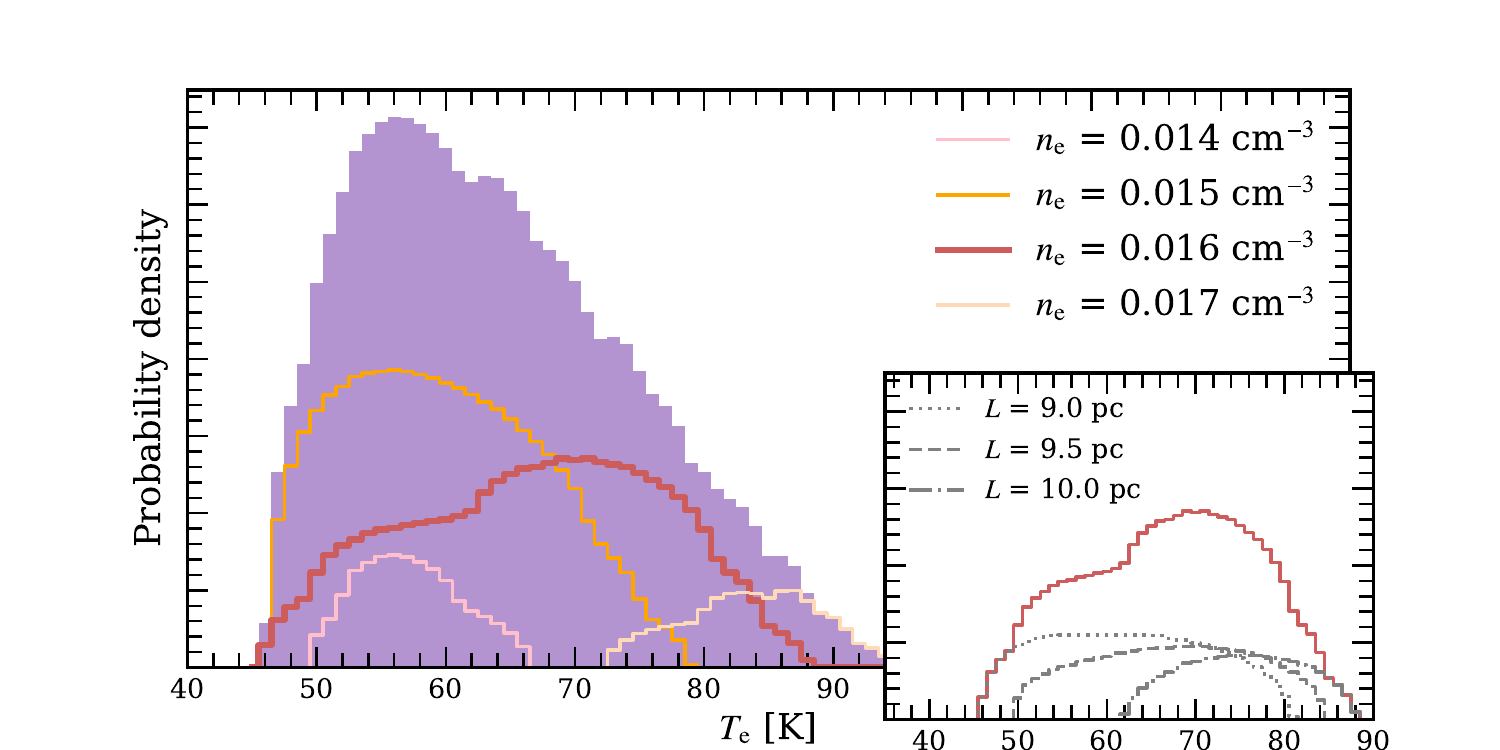}
    \caption{Illustration of the influence of the parameters $n_\mathrm{e}$ and $L$ on the optimal value found for $T_\mathrm{e}$, for Cygnus A. The purple histogram represents all the values of $T_\mathrm{e}$ that gives $\chi^2_\mathrm{r}(T_\mathrm{e}) -\chi^2_{\rm r, opt} < 30\% \times \chi^2_{\rm r, opt}$. The coloured curves represents the values of $T_\mathrm{e}$ that verify the condition while $n_\mathrm{e}$ is fixed at a given value. The snapshot on the right panel is a focus on the curve with $n_\mathrm{e} = \SI{0.016}{\per\centi\meter\cubed}$. The gray curves represent the values of $T_\mathrm{e}$ that verify $\chi^2_\mathrm{r}(T_\mathrm{e},n_\mathrm{e}=\SI{0.016}{\cmcube}) -\chi^2_{\rm r, opt} < 30\% \times \chi^2_{\rm r, opt}$ while $L$ is fixed at given values. }

    \label{fig:multimodality}
\end{figure}

In this section we discuss the structure of the 5-dimensional physical parameter space (as illustrated in Figs. \ref{fig:casa-47-histo} to \ref{fig:cyga-histo}). We also present the impact of multimodality on the evaluation of physical constraints.
The best-constrained parameter in the 5D parameter space is the turbulent velocity. The optimal value of $\varv_\mathrm{t}$ does not depend on the local minimum we look at and is very stable throughout the grid search.
{As shown in Fig. \ref{fig:multimodality},} for a given $n_\mathrm{e}$, the probability density for $T_{\rm e}$ forms a bell curve around a $T_\mathrm{e, opt}$. The overall structure for the parameter $T_\mathrm{e}$ is thus a combination of bell curves, one for each different $n_\mathrm{e}$ comprised in the 30\% uncertainty limit. We generally find few solutions for $n_\mathrm{e}$, and the different bell curves for $T_\mathrm{e}$ are relatively close to each other, leading to a total uncertainty of less than 10~K in some cases.
The values chosen for $T_0$ and $L$ seem to influence the optimal values of $n_\mathrm{e}$ and $T_\mathrm{e}$ and not the general structure of the solutions. The parameter $L$ impacts the value of $n_\mathrm{e}$ and hence also the positions of the curve bells of $T_\mathrm{e}$ (see Fig.~\ref{fig:multimodality}), whereas $T_0$ seems to lightly influence the position peak of $\varv_\mathrm{t}$, resulting in a more or less spread bell curve for $\varv_\mathrm{t}$.

The complex nature of the parameter space we study, as well as the intrinsic degeneracy of the optimisation problem, make it difficult to conduct a proper minimization methodology. 
We are currently looking to improve this aspect of our analysis, which is a work in progress that is beyond the scope of this article. 

\section{Concluding remarks}
\label{sec:conc}

We have shown that NenuFAR can be used to quantitatively study Radio Recombination Lines in the lowest observable frequency range, in order to enhance our understanding of the diffuse ISM. More specifically, we have used C$\alpha$RRLs as a diagnostic tool to provide constraints on the electron density $n_\mathrm{e}$ and electron temperature $T_\mathrm{e}$ in four different diffuse clouds of the ISM. 
Using the NenuFAR telescope's low-frequency capabilities, we conducted observations of the two brightest radio sources of the northern sky: Cassiopeia A and Cygnus A. Through a meticulous data processing pipeline, we removed radio frequency interference and enhanced the S/N, resulting in high-quality spectral line profiles. This process allowed us to detect numerous carbon recombination lines, with Cassiopeia A yielding 398 lines, significantly improving upon previous studies. The lines towards Cygnus A lines are fainter and require extensive line stacking. Upon stacking, we obtained 11 usable lines towards Cyg A and 28 towards Cas A. 

NenuFAR has significantly advanced the study of carbon Radio Recombination Lines (CRRLs) by providing unprecedented sensitivity (of $\sim$2~Jy at 50~MHz, for a 2-hour integrating time at  resolution) and broader frequency coverage in the 10–85 MHz range, at an unprecedented spectral resolution of 95~Hz. This work demonstrates NenuFAR’s ability to resolve finer details in CRRL profiles, achieving an improved S/N by a factor of 4 to 10 compared to similar observations with LOFAR. The study addressed blending challenges and discrepancies in line intensities through meticulous data reduction and analysis, establishing a new benchmark for low-frequency radio observations. These advancements provided refined measurements of physical properties, such as electron density and temperature, in line-of-sight clouds. 

Furthermore, the methodology developed includes robust RFI mitigation and adaptive line-stacking techniques. It sets a precedent for future studies in diffuse ISM characterization. Future improvements will focus on automating parts of the data pipeline and extending observations to additional sources. Overall, this study positions NenuFAR as a transformative tool in low-frequency radio astronomy, paving the way for more comprehensive studies of Galactic and extragalactic CRRLs. 

We inferred physical parameters for the four ISM clouds detected towards Cas A and Cyg A. 
The first cloud is detected towards Cas A at a central LSR velocity of -47~km~s$^{-1}$. Our best-fit model is achieved for $T_{\rm e}=45$~K, $n_{\rm e}=0.027~\cmcube$, $T_0=1200$~K, $L=33.0$~pc and $\varv_t=3.3$~km~s$^{-1}$. From these we derive EM$_\ion{C}{II} = \SI{0.024}{\per\centi\meter^{-6}~pc}$, $p_{\rm th}/k = \SI{8.7e3}{K.\cmcube}$ and $p_{\rm t}/k = \SI{3.4e5}{K.\cmcube}$. 
The second cloud is detected towards Cas A at a central LSR velocity of -38~km~s$^{-1}$. Our best-fit model is achieved for $T_{\rm e}=40$~K, $n_{\rm e}=0.025~\cmcube$, $T_0=1200$~K, $L=20.5$~pc and $\varv_t=4.7$~km~s$^{-1}$. From these we derive EM$_\ion{C}{II} = \SI{0.013}{\per\centi\meter^{-6}~pc}$, $p_{\rm th}/k = \SI{7.1e3}{K.\cmcube}$ and $p_{\rm t}/k = \SI{6.4e5}{K.\cmcube}$.
The third cloud is detected towards Cas A at a LSR velocity of 0~km~s$^{-1}$. Our best-fit model is achieved for $T_{\rm e}=52$~K, $n_{\rm e}=0.014~\cmcube$, $T_0=1000$~K, $L=15.0$~pc and $\varv_t=3.7$~km~s$^{-1}$. From these we derive EM$_\ion{C}{II} = \SI{0.003}{\per\centi\meter^{-6}~pc}$, $p_{\rm th}/k = \SI{5.2e3}{K.\cmcube}$ and $p_{\rm t}/k = \SI{2.2e5}{K.\cmcube}$.
The fourth cloud is detected towards Cyg A at a LSR velocity of 3.5~km~s$^{-1}$. Our best-fit model is achieved for $T_{\rm e}=77$~K, $n_{\rm e}=0.018~\cmcube$, $T_0=1600$~K, $L=8$~pc and $\varv_t=7.9$~km~s$^{-1}$. From these we derive EM$_\ion{C}{II} = \SI{0.003}{\per\centi\meter^{-6}~pc}$, $p_{\rm th}/k = \SI{9.9e3}{K.\cmcube}$ and $p_{\rm t}/k = \SI{13.0e5}{K.\cmcube}$.
These results are generally compatible with previous LOFAR studies, and present typical values for diffuse partially ionised ISM phases. 

The complexity induced by line blending and the intrinsic degeneracy of the physical modelling highlights the importance of using different observational techniques on parameter estimation. 

Discrepancies observed between LOFAR and NenuFAR can be attributed to differences in beam size and field of view, which may lead to increased turbulent velocities and/or a dilution effect, particularly impacting the line intensity seen in Cas A. 
The turbulent velocity emerged as the most stable parameter throughout our analysis. For every cloud, the probability distribution of the electron temperature exhibited a bell curve influenced by the input values for $T_0$ and $L$.

\begin{acknowledgements}
This work made use of Astropy (http://www.astropy.org): a community-developed core Python package and an ecosystem of tools and resources for astronomy \citep{AstropyCollaboration13, AstropyCollaboration18, AstropyCollaboration22}. The authors acknowledge the support of the Programme National \lq Physique et Chimie du Milieu Interstellaire' (PCMI) of CNRS/INSU with INC/INP co-funded by CEA and CNES. This paper is based on data obtained using the NenuFAR radiotelescope. NenuFAR has benefited from the funding from CNRS/INSU, Observatoire de Paris, Observatoire Radioastronomique de Nançay, Observatoire des Sciences de l’Univers de la R\'egion Centre, R\'egion Centre-Val de Loire, Universit\'e d’Orl\'eans, DIM-ACAV and DIM-ACAV+ de la R\'egion Ile de France, and Agence Nationale de la Recherche. We acknowledge the collective work from the NenuFAR-France collaboration for making NenuFAR operational, and the Nançay Data Center resources used for data reduction and storage. This project has received funding from the French Agence Nationale de la Recherche (ANR) through the projects COSMHIC (ANR-20-CE31-0009) and COSMHIC\_Ukraine (ANR-20-CE31-0009-03). AB acknowledges financial support from the INAF initiative \lq IAF Astronomy Fellowships in Italy'. J.R.G. thanks the Spanish MCINN for funding support under grant PID2023-146667NB-I00.

\end{acknowledgements}

\bibliographystyle{aa}
\bibliography{biblio}
\onecolumn
\begin{appendix}

\section{Discarding anomalous subbands in the correction of a given subband}
\label{sec:dasitcoags}

We denote by $S_\mathrm{i} (\nu)$ the intensity in the $i$-th subband we want to correct: the set of subband used to compute the average is $S_{\rm i-5}$ to $S_{\rm i+5}$. So as to not spread RFI or distortions of the baseline throughout the 11 subbands, we flag those who differ from the subband we want to clean. To determine if a subband from the set $\{S_{\rm k}\}_{\rm k\in[i-5,i+5]}$ is significantly different from $S_\mathrm{i}$, we apply a statistical criterion: the level of distortion of $S_\mathrm{i}$ is assessed by computing the 2-norm of $S_\mathrm{i}$, defined by the square root of the integral of the squared subband:
\begin{equation}
    \| S_\mathrm{i}\|_2 = \sqrt{\int S_\mathrm{i}^2(\nu) \rm d}\nu.
\end{equation}

We then assess the level of dissimilarity of each $(S_\mathrm{k})_{\rm k\in[i-5,i+5]}$ to $S_\mathrm{i}$ by computing the square of the 2-distance between $S_\mathrm{k}$ and $S_\mathrm{i}$:
\begin{equation}
    \forall k \in [i-5, i+5], \ k\neq i, \ \| S_\mathrm{k} - S_\mathrm{i} \|_2^2 = \int (S_\mathrm{k} - S_\mathrm{i})^2 \mathrm{d}\nu
\end{equation}
We remove every subband whose dissimilarity level is above 5\% of the distortion level of the original subband. This criterion was chosen empirically. As more distorted subbands tend to have neighbours also presenting distortion, it can happen that a median subband baseline cannot be generated this way. In this case, we simply flag out the $S_{\rm i}$ subband. The resulting potential baseline deformation are dealt with in a subsequent flattening step. The median of the subbands that have not been flagged is smoothed using a Savitzky-Golay filter (dotted black line in the third panel of Fig.~\ref{fig:reduc-subband-scale}), and subtracted from $S_\mathrm{i}$. We call $S_\mathrm{i}'$ the result of the subtraction (blue line in the fourth panel of Fig.~\ref{fig:reduc-subband-scale}).

\section{Theory of line profiles and line fitting}
\label{sec:fitlin}

The shape of low frequency radio recombination lines can be used as a probe of the physical conditions of the diffuse ISM in which they are produced. In Sect. \ref{sub:line-theory}, we present the theory of RRLs and to what extent they relate to the physical parameters of the ISM. In Sect. \ref{sub:line-fitting} we describe how we used this theory to constrain the physical parameters of ISM in the sightline of our specific sources.

\subsection{Line profile: Theory}
\label{sub:line-theory}

In the diffuse ISM, the recombination of ions and electrons forms atoms whose electrons are in levels with high principal quantum numbers ($n$). These newly bound electrons decay from level to level, radiating photons in form of spectral lines in the process. When these lines appear in the radio range, so-called \lq Radio recombination lines' are generated, that is after the de-excitation of an atom in a Rydberg state from the electronic quantum state $n$ to a lower electronic quantum state $n'$. Here, we study carbon radio recombination lines and we focus on the $\alpha$ transition (i.e. from $n+1$ to $n$) with $n \in [400,850]$. As described in \citet{Salgado17b}, for this range of $n$ and for the physical conditions of CNM, radio recombination lines are seen in absorption.
The frequency of a recombination line is given by e.g. \cite{Gordon02}:
\begin{equation}
    \nu_{\rm n+1\rightarrow n} = R_\infty c \frac{m_\mathrm{H} - m_\mathrm{e}}{m_\mathrm{H}} \left( \frac{1}{n^2} - \frac{1}{(n+1)^2}\right) 
\end{equation}
with $R_\infty = \SI{109732.30}{\per\centi\meter}$ (the Rydberg constant for the carbon atom) and $m_\mathrm{H}$ and $m_\mathrm{e}$ the masses of the hydrogen atom and the electron. 

We use a Voigt profile to model the shape of the line, as in \citet{Gordon02}, which is described by three parameters, namely the integrated optical depth and the Gaussian and Lorentzian width{s}. In turn, these parameters depend on the physical conditions of the diffuse ISM that we want to constrain. In this section, we detail the equations used to infer the physical conditions of diffuse ISM clouds from the shape of recombination lines.

\paragraph{Voigt profile.} A Voigt profile is the convolution product of a Lorentzian profile and a Gaussian (Doppler) profile. We performed the fitting using the \texttt{astropy.modelling.Voigt1D} function from the \texttt{astropy} project\footnote{\href{https://www.astropy.org}{https://www.astropy.org}}, in which the Voigt profile is described by four independent parameters: its central velocity $\varv_0$, the amplitude of the Lorentzian $a_\mathcal{L}$, the FWHM of the Gaussian $w_\mathcal{D}$ and the FWHM of the Lorentzian $w_\mathcal{L}$. Let $\mathcal{V}$ be the Voigt profile function:
\begin{equation}
    \mathcal{V}_{\varv_0, \, a_\mathcal{L}, w_\mathcal{L}, w_\mathcal{D}}(\varv) = \int\limits_{-\infty}^{+\infty} \mathcal{G}_0(\varv;\varv_0=0,w_\mathcal{D}) \, \mathcal{L}(\varv';\varv_0,w_\mathcal{L},a_\mathcal{L}) \, \rm d\varv', \forall \varv
\end{equation}
with $\mathcal{G}_0$ the normalised Gaussian profile and $\mathcal{L}$ the Lorentzian of given amplitude. In practice, the $\varv_0$ value is set by the observations, leaving the Voigt profile to depend only on $(a_\mathcal{L}, w_\mathcal{L}, w_\mathcal{D})$. The FWHM of the Voigt profile $w_\mathcal{V}$ depends on those of the Lorentzian and Gaussian ones, following \citet{Gordon02}:
\begin{equation}
w_\mathcal{V}(w_\mathcal{L},w_\mathcal{D}) = (0.5346 \, w_\mathcal{L} + \sqrt{0.2166w_\mathcal{L}^2 + w_\mathcal{D}^2}).
\end{equation}

\paragraph{Integrated optical depth.} For a line $n \rightarrow n'$, \citet{Salgado17a} provides the following description of the integrated line-to-continuum intensity:
\begin{equation}
\begin{array}{cl}
    A_\mathrm{n}& \equiv \displaystyle\int\frac{I_\mathrm{n}^{line}(\nu)}{I_\mathrm{n}^{cont}(\nu)}d\nu \\ 
    &= \displaystyle -0.2 \, b_\mathrm{n} \, \beta_{\rm n,n'} \, \left( \frac{T_\mathrm{e}}{\SI{100}{\kelvin}} \right)^{-2.5}\left( \frac{n_\mathrm{e}}{ \SI{0.1}{\per\centi\meter\cubed}} \right)^{2} \left( \frac{L}{\si{pc}} \right) \si{\hertz}. 
\end{array}
\end{equation}
In this formula, $n_{\rm e}$ and $T_{\rm e}$ are the local density and temperature of electrons. $L$ is the size of the absorbing cloud. $b_\mathrm{n}$ is a departure coefficient from local thermodynamic equilibrium (LTE) that depends on $T_\mathrm{e}$ and $n_\mathrm{e}$, and $\beta_{\rm n,n'}$ is a correction factor for stimulated emission that has to be accounted for in the equations of statistical equilibrium; it also depends on $T_\mathrm{e}$ and $n_\mathrm{e}$. The equation (7) from \citet{Salgado17a} also accounts for a dependency on the local radiation field temperature $T_0$, however this dependency is negligible for $n> 200$ (see \cite{Prozesky18}). Figure (8) from \citet{Oonk17} shows from observations of Cas A that varying $T_0$ has little effect on the integrated area of the lines. For $\alpha$ transitions, the $\beta_{\rm n,n+1}$ factors are tabulated over a sample grid of $T_\mathrm{e}$ and $n_\mathrm{e}$, with a constant $T_0=0$~K \citep{Salgado17a}. 

\paragraph{Amplitude of the Lorentzian.} The amplitude of the Lorentzian can be inferred from the line integrated intensity $A_\mathrm{n}$  and the FWHM of the Lorentzian and Gaussian profiles \citep{Gordon02}:
\begin{equation}
\begin{array}{cl}

    a_\mathcal{L}(A_{\rm n},w_\mathcal{L},w_\mathcal{D}) & = \displaystyle \frac{A_{\rm n}}{p(w_\mathcal{L},w_\mathcal{D})\times w_\mathcal{V}(w_\mathcal{L},w_\mathcal{D})} \vspace{2mm} \\
    \text{with } p(w_\mathcal{L},w_\mathcal{D})&=(1.57 - 0.507 \exp{[-0.85 w_\mathcal{L}/w_\mathcal{D}]}). \vspace{2mm}\\
    
\end{array}
\end{equation}

\paragraph{Full width half maximum of the Lorentzian.} Three distinct physical effects control the width of the Lorentzian component $w_\mathcal{L}$: radiation broadening, natural broadening, and pressure (or collisional) broadening. Each of them is characterized by a width $w_\mathrm{rad}$, $w_\mathrm{nat}$, $w_\mathrm{col}$ such that $w_\mathcal{L} = w_{\rm rad} + w_{\rm nat} + w_{\rm col}$. 
The radiation broadening can be modelled by: 
     \begin{equation}
    \begin{array}{cl}
w_{\rm rad}& = \displaystyle\frac{2}{\pi} \times 2.137 \ 10^{-4} \times \left(\frac{2 \mathrm{R_\infty c}}{\nu_{0,\alpha}}\right)^{\alpha+1} \times \mathrm{k_B} T_0 \, \nu_{0, \alpha} \, n^{-3\alpha -2} \\
     & \hspace{4.55cm} \times \sum\limits_{\Delta n = 1}^{\infty}(\Delta n)^{\alpha - 2} \si{\hertz}; 
    \end{array}
    \end{equation}

In this expression, $T_{\rm 0}$ is the brightness temperature of the ambient radiation field $T_b(\nu)$ at a given frequency $\nu_{\rm 0, \alpha}$, representative of the observation range. $\alpha$ is then the spectral index of the radiation field variation in the same frequency range, such that:

\begin{equation}
    T_b(\nu) = T_0 \left(\frac{\nu}{\nu_{0,\alpha}} \right)^\alpha.
\end{equation}
We chose $\alpha=-2.6$ and $\nu_{0,\alpha} = 100$~MHz, consistently with \citet{Salgado17a} and \citet{Salas17}. In addition to this, the natural broadening depends only on $n$ through: 
\begin{equation}
w_{\rm nat} = 1.2 \times 10^{-6} \,\nu_{\rm n+1\rightarrow n} \, \displaystyle\frac{\ln{n}}{n^2} \si{\hertz}. 
\end{equation}
Finally, the pressure (or collisional) broadening depends on $T_\mathrm{e}$ and $n_\mathrm{e}$:
\begin{equation}
     w_{\rm col} = \frac{1}{2\pi}n_\mathrm{e} 10^{\alpha(T_\mathrm{e})} n^{\gamma(T_\mathrm{e})} \si{\hertz}.
\end{equation} 
where $\alpha$ and $\gamma$ are tabulated in \citet{Salgado17b}.

\paragraph{Full width half maximum of the Gaussian: Doppler broadening.} The Gaussian shape in the Voigt profile is physically due to the Doppler broadening. The Doppler effect in diffuse clouds originates in the combined action of the thermal and non-thermal motions. The non-thermal motions include disordered motions, as well as potential ordered-motions such as infall, rotation, or shocks. To simplify, we will refer to all the non-thermal motions as \lq turbulence'.
Thus, the FWHM of the Gaussian is the quadratic sum of these two effects \citep{Salgado17b}:
\begin{equation}
    w_\mathcal{D}(T_\mathrm{e}, \varv_\mathrm{t}) = \frac{\sqrt{4\ln{2}}\nu_{\rm n+1\rightarrow n}}{c} \times \sqrt{ \frac{2k_BT_\mathrm{e}}{M_\mathrm{C}} + \varv_\mathrm{t}^2 }
    \label{eq:doppler-width}
\end{equation}
with $M_C$ the atomic mass of carbon. The average turbulent velocity $\varv_t$ is the mean of the turbulent motions within the telescope beam with respect to the angular extent of the observed cloud. Hence, it depends greatly on the beamwidth of the instrument as well as the scale of the ISM phase along the line of sight.

\paragraph{Dependencies.} Overall, the dependencies of the observables (\textit{i.e.,} the parameters of the line shape: $A_\mathrm{n}$, $w_{\mathcal{L}, \mathrm{n}}$, $w_{\mathcal{D}, \mathrm{n}}$, $a_{\mathcal{L}, \mathrm{n}}$) on the physical parameters ( $T_{\rm e}$, $n_{\rm e}$, $\varv_{\rm t}$, $T_0$ and $L$) can hence be summarized as:
\begin{equation}
    \begin{array}{cl}
        A_\mathrm{n} &\equiv A_\mathrm{n}(T_\mathrm{e}, n_\mathrm{e}, L) \\
        w_{\mathcal{L}, \mathrm{n}} &\equiv w_{\mathcal{L}, \mathrm{n}} (T_\mathrm{e}, n_\mathrm{e}, T_0) \vspace{2mm}\\
        w_{\mathcal{D}, \mathrm{n}} &\equiv w_{\mathcal{D}, \mathrm{n}} (T_\mathrm{e}, \varv_\mathrm{t}) \vspace{2mm}\\ 
        a_{\mathcal{L}, \mathrm{n}} &\equiv a_{\mathcal{L}, \mathrm{n}}(A_\mathrm{n}, w_{\mathcal{L},\mathrm{n}}, w_{\mathcal{D},\mathrm{n}})\\
        & \equiv a_{\mathcal{L}, \mathrm{n}}(T_\mathrm{e}, n_\mathrm{e}, L, \varv_\mathrm{t}, T_0)
        
    \end{array}
    \label{eq:dependancies}
\end{equation}

\paragraph{Additional diagnostics.} From these physical parameters, we are able to infer additional diagnostic tools, namely the emission measure of \ion{C}{II} EM$_\ion{C}{II}$, the thermal pressure $p_{\rm th}$ and the turbulent pressure $p_{\rm t}$. 

As the ionisation energy of carbon is lower than for hydrogen and helium (11.26~eV, 13.6~eV and 24.59~eV respectively), we assume carbon to be the main ionised species in the observed PDRs. Hence we have $n_\ion{C}{II} \simeq n_{\rm e}$, and EM$_\ion{C}{II}$ can be inferred from the measured parameters $n_{\rm e}$ and $L$ (see Eq. \ref{eq:EMC+}).
\begin{equation}
    EM_{\ion{C}{II}} = n_{\rm e} \times n_\ion{C}{II}\times  L = n_{\rm e}^2 \times L
    \label{eq:EMC+}
\end{equation}
We also compute the thermal pressure $p_{\rm th}$ of the observed clouds. We assume that electrons are at thermal equilibrium, hence $T_{\rm e} = T$. As done in \citet{Oonk17}, we assume that carbon is fully ionised, and the \ion{C}{II} fraction is $x(\ion{C}{II})\simeq 1.4 \times10^{-4}$ in diffuse lines of sights \citep{Sofia04}. 
\begin{equation}
    p_{\rm th}/k = n_{\rm H}\times T = \frac{n_{\rm e}}{x(\ion{C}{II})} \times T_{\rm e}
\end{equation}

Finally, we compute the turbulent pressure $p_{\rm t}$ from the measured $\varv_{\rm t}$, using the equation from \citet{Lequeux02Chap13}:
\begin{equation}
    p_{\rm t} = \frac{1}{3}\rho \varv_{\rm t}^2 = \frac{1}{3} n_{\rm H}m_{\rm H} \varv_{\rm t}^2 = \frac{n_{\rm e}m_{\rm H}}{3x(\ion{C}{II})} m_{\rm H} \varv_{\rm t}^2,
\end{equation}
with $m_{\rm H}$ the mass of the hydrogen atom. Thus, we have:
\begin{equation}
    p_{\rm t}/k = \frac{ n_{\rm e} \ m_{\rm H} \ \varv_{\rm t}^2}{3 \ k \ x(\ion{C}{II})}
\end{equation}

For the neutral phase of the Milky Way, \citep{Wolfire03} predicts that the turbulent pressure dominates over the thermal pressure, which is typically of $\sim\SI{10e3}{\kelvin\cmcube}$.

\subsection{Line profile: Fitting}
\label{sub:line-fitting}
For each source, we began by fitting individually the stacked lines introduced in Sects~\ref{sub:casao} and \ref{sub:cygao} with Voigt profiles. We then determined the profile broadening and integrated intensity. We subsequently studied the variation of these two parameters as a function of the average principal quantum number of the transitions present in the stacks $n$. This average quantum number is the one that is used to compute the departure coefficients as well as the radiative, natural and collisional widths. Finally, we ran through a grid of physical parameters to find which combination of defining parameters $(T_\mathrm{e}$ the electron temperature, $n_\mathrm{e}$ the electron density, $T_0$ the temperature of the local radiation field, $\varv_\mathrm{t}$ the local mean turbulent velocity and $L$ the depth of the cloud along the line-of-sight) best reproduces the observed variation as a function of $n$. 

\subsubsection{Fitting the shape of the lines}
\label{subsub:fitting-shape-line}
The algorithm iterates over the individual spectral stacks obtained from observations. Each iteration involves:
\begin{enumerate}

    \item Signal detection: we calculate a S/N threshold, and the algorithm determines whether a significant signal is present in the data. We deem the signal significant if the S/N is above 3. To compute the signal value, we integrate the spectral line. To compute the noise value, we mask the line out of the spectral stack, and we compute the rms of the remaining frequency channels. We then integrate this value over the frequency span of the spectral stack.
    
    \item Voigt profile fitting: if a signal is detected, the algorithm fits a Voigt profile to the spectral line using the curve-fitting function (\texttt{curve\_fit}) of the \texttt{scipy.optimize} module from \texttt{scipy}\footnote{https://scipy.org} \citep{Virtanen20}. Generally the Voigt fitting step does not require any prior knowledge of the signal. However, for lines of sight with blended velocity components, constraints can be added to guide the algorithm through the curve-fitting (as for Cas A, see Sect.~\ref{subsub:casao-linefitting}).
    \item Parameter extraction: line width and integrated intensity are extracted from the fitted Voigt profile. The fitting uncertainties are determined by integrating the residuals of the fit. Again, for lines of sight with blended velocity components, we had to adopt a different procedure (see Sect.~\ref{subsub:casao-linefitting} for Cas A). 
\end{enumerate}

\subsubsection{Extracting the physical parameters}
\label{subsub:phys-param-extraction}

We computed the $\chi^2$ distance between a model and the observation based on line width and integrated intensity. For a given target, we call $n_{\rm S}$ the number of stacks for which we extracted these two quantities. Let $A$ be the list of observed integrated intensities, $\widehat{A}$ the list of modelled integrated intensities, and $\sigma_A^2$ the variance on the measure of $A$, as defined in Sect. \ref{subsub:fitting-shape-line}.
Let $w_\mathcal{V}$ be the list of observed profile widths, $\widehat{w_\mathcal{V}}$ the list of modelled profile widths, and $\sigma_{w_\mathcal{V}}^2$ the variance on the measure of $w_\mathcal{V}$. The $\chi^2$ distance is:
\begin{equation}
    \chi^2 = \sum\limits_{k=1}^{n_\mathrm{S}} \frac{(\widehat{w_{\mathcal{V},k}}-w_{\mathcal{V},k})^2}{\sigma_{w_\mathcal{V},k}^2} + \sum\limits_{k=1}^{n_\mathrm{S}} \frac{(\widehat{A_k}-A_k)^2}{\sigma_{A,k}^2}.
    \label{eq:chi2}
\end{equation} We then evaluated the reduced $\chi^2$, noted $\chi_r^2$, by dividing by the number of degrees of freedom of the optimisation problem:
\begin{equation}
    \chi^2_r = \frac{\chi^2}{2 \, n_\mathrm{S} - n_\mathrm{P}},
    \label{eq:chi2_reduced}
\end{equation}
with $n_\mathrm{P}=5$ the number of independent parameters of the physical modelling. We used 2$\times n_\mathrm{S}$ for the total number of data points as we extracted two observables from each velocity component in a stack ($w_\mathcal{V}$ and $A$). 

The parameters $(T_\mathrm{e}, n_\mathrm{e}, T_0, \varv_\mathrm{t})$ are systematically explored within predefined linear ranges chosen to be representative of the CNM that we initially sought after. The $L$ parameter is explored around a prior determined by a cloud identification method, described in Sect.~\ref{subsub:cloud-id}. The range and coarseness of the initial grid are in Table \ref{tab:param_precision}.
\begin{table}
    \centering
    \caption{Linear ranges and coarseness of the grids explored to infer physical parameters.}
    \begin{tabular}{|c|c|c|c||c|}
    \hline
        Parameter & Start & Stop & Step &  Final precision \\ \hline 
        $T_\mathrm{e}$  [K] & 10 & 5000 & 10 & 1 \\
        $n_\mathrm{e}$  [\si{\per\centi\meter\cubed}] & 0.01 & 1 & 0.01 & 0.001 \\
        $T_0$           [K] & 100 & 5000 & 500 & 100 \\
        $\varv_\mathrm{t}$ [km~s$^{-1}$] & 1 & 20 & 1 &  0.1 \\
        $L$ [pc] & $L_{\rm min}$\tablefootmark{(a)} & $L_{\rm max}$\tablefootmark{(a)} & 1 & 0.5 \\ \hline
    \end{tabular}
    \tablefoot{Left part: parameters of the broad initial grid, the columns start and stop define the range and column step defines the coarseness. 
    Last column: precision of the finest grid. \\
    \tablefoottext{a}{{The edges of the exploratory range are determined through cloud identification as $L_{\rm min} = L_{\rm prior} - 10$~pc and $L_{\rm max} = L_{\rm prior} + 10$~pc
    }}}
    \label{tab:param_precision}
\end{table}

Exploration of the parameter grid revealed that the modelling does not provide significant constraints on $L$ and $T_0$, leading to the existence of multiple local minima. The preliminary explorations show that $\varv_\mathrm{t}$ is the same in every local minimum, whereas the values found for $T_\mathrm{e}$ and $n_\mathrm{e}$ are shifting when varying $L$ or $T_0$. In order to lift part of the degeneracy, we chose to restrict the exploration of the finer parameter grid to values close to the global minimum of $T_0$ and $L$ obtained during the broad exploration. 
For each combination of parameters within the pre-defined range, model predictions are compared against observed data using $\chi_r^2$. In a second step, we refined the grid around the detected minima. The maximum precision to which the grid is refined for each parameter is detailed in Table \ref{tab:param_precision}.

\subsubsection{Cloud identification}
\label{subsub:cloud-id}
We performed a cross-analysis of CRRLs from NenuFAR, dust clouds in a radius of 1.25~kpc from the Sun, and CO PPV cubes towards each source in order to locate the observed clouds along our lines-of-sight.
The 3D dust maps are from the Gaia DR3 \citep{Edenhofer24}, and CO data cubes are from the whole-Galaxy CO (1 $-$ 0) survey from the 1.2m Millimeter-Wave telescope at the Center for Astrophysics | Harvard \& Smithsonian \citep{Dame01}. This survey is a composite map with varying specificities. The map containing Cas A yields a spectral resolution of 0.65~km~s$^{-1}$ and and an angular resolution of 7.5' meanwhile the map containing Cyg A has the same spectral resolution and an angular resolution of 15'.
The cloud identification provides a meaningful prior for the grid search for the $L$ parameter, as there is no \textit{a priori} value for the CNM.

First, we searched in the Gaia DR3 maps if dust clouds were present in our lines of sight. For each of them, we found a number of clouds encompassed (totally or partially) in the beam of NenuFAR. The 3D nature of the map allowed us to measure the length of the dust clouds by measuring the full width half maximum (FWHM) of their absorption along the line of sight (see top panels of Figs. \ref{fig:cloud-identification-0},\ref{fig:cloud-identification-4738}). It also allowed us to assess the structure of the clouds by projecting the 3D map on the plane of sky. Combining these informations, we were able to extract the structures in the plane of sky of each absorbing cloud along the line-of-sight separately, by projecting only the slices of the 3D absorption map that contained a given cloud. Then, for each source, we computed the CO spectrum averaged over the NenuFAR beam to identify common velocity components between CRRLs and CO. If we found CO emission at the same central velocity as the lines we detected in CRRLs, then we selected a slice of the CO cube containing the emission (see top panels of Figs. \ref{fig:cloud-identification-0},\ref{fig:cloud-identification-4738}), and we computed its 0$^{\rm th}$ order moment so as to assess the size and shape of a potential CO cloud at this velocity. Finally, we overlaid the projected maps of the dust clouds in the plane-of-sky with the $0^{\rm th}$ order moment in CO for different velocities, to compare the shapes of dust and CO clouds, and find potential matches (see bottom panels of Figs.\ref{fig:cloud-identification-0},\ref{fig:cloud-identification-4738}).

In absence of such a match, we compute the expected local standard of rest (LSR) velocity as a function of the heliocentric distance, based on the theoretical modelling of \citet{Reid19} and the python module of \citet{Wenger18}. The methodology of \citet{Reid19} is to first assume a circular galactocentric velocity, and to project the velocity vector on the heliocentric line-of-sight according to the galactic longitude and latitude of the source, as well as the Sun's proper motion. Then, we check if a dust cloud in the beam whose LSR velocity is the one we detect with CRRLs.

The prior on $L$ we obtain through this method is rudimentary because CO, diffuse and dusty clouds are not necessarily correlated. However, this method allowed us to identify a prior for every cloud, which our grid explored $\pm$10 pc around the prior in steps of 1~pc. 

\section{Detections and line fitting}

\begin{table*}[h]
\centering
\caption{Summary of the parameters of the stacking procedure for each source.}
\begin{tabular}{|c|cccc|}
\hline
Source                  & Frequency range & \makecell{{Average} quantum \\ number} & \makecell{Number of lines \\ in the preselection} & \makecell{Effective number of \\ lines in the stack} \\ \hline
\multirow{28}{*}{Cas A} & 79.11 - 84.41   & 431                              & 10                                  & 7                                      \\
                        & 73.91 - 79.11   & 441                              & 10                                  & 7                                      \\
                        & 69.16 - 73.91   & 451                              & 10                                  & 10                                     \\
                        & 64.81 - 69.16   & 461                              & 10                                  & 7                                      \\
                        & 60.81 - 64.81   & 471                              & 10                                  & 8                                      \\
                        & 57.14 - 60.81   & 481                              & 10                                  & 8                                      \\
                        & 53.76 - 57.14   & 491                              & 10                                  & 8                                      \\
                        & 50.63 - 53.76   & 501                              & 10                                  & 8                                      \\
                        & 47.75 - 50.63   & 511                              & 10                                  & 8                                      \\
                        & 45.08 - 47.75   & 521                              & 10                                  & 7                                      \\
                        & 42.61 - 45.08   & 531                              & 10                                  & 8                                      \\
                        & 40.31 - 42.61   & 541                              & 10                                  & 9                                      \\
                        & 38.18 - 40.31   & 551                              & 10                                  & 9                                      \\
                        & 36.19 - 38.18   & 561                              & 10                                  & 6                                      \\
                        & 34.34 - 36.19   & 571                              & 10                                  & 8                                      \\
                        & 32.61 - 34.34   & 581                              & 10                                  & 7                                      \\
                        & 31.00 - 32.61   & 591                              & 10                                  & 8                                      \\
                        & 29.49 - 31.00   & 601                              & 10                                  & 8                                      \\
                        & 26.76 - 29.49   & 616                              & 20                                  & 19                                     \\
                        & 24.35 - 26.76   & 636                              & 20                                  & 18                                     \\
                        & 22.22 - 24.35   & 656                              & 20                                  & 17                                     \\
                        & 20.34 - 22.22   & 676                              & 20                                  & 17                                     \\
                        & 18.66 - 20.34   & 696                              & 20                                  & 16                                     \\
                        & 16.47 - 18.66   & 721                              & 30                                  & 24                                     \\
                        & 14.61 - 16.47   & 751                              & 30                                  & 27                                     \\
                        & 13.02 - 14.61   & 781                              & 30                                  & 29                                     \\
                        & 11.65 - 13.02   & 811                              & 30                                  & 30                                     \\
                        & 10.47 - 11.65   & 841                              & 30                                  & 29                                     \\ \hline
\multirow{11}{*}{Cyg A} & 69.16 - 84.41   & 441                              & 30                                  & 30                                     \\
                        & 57.14 - 69.16   & 471                              & 30                                  & 30                                     \\
                        & 47.75 - 57.14   & 501                              & 30                                  & 30                                     \\
                        & 40.31 - 47.75   & 531                              & 30                                  & 27                                     \\
                        & 34.34 - 40.31   & 561                              & 30                                  & 22                                     \\
                        & 29.49 - 34.34   & 591                              & 30                                  & 26                                     \\
                        & 24.35  - 29.49  & 626                              & 40                                  & 38                                     \\
                        & 20.34 - 24.35   & 666                              & 40                                  & 34                                     \\
                        & 17.16 - 20.34   & 706                              & 40                                  & 35                                     \\
                        & 14.05 - 17.16   & 751                              & 50                                  & 36                                     \\
                        & 11.65 - 14.05   & 801                              & 50                                  & 44                                     \\ \hline

\end{tabular}

\label{tab:stacks}
\end{table*}

\begin{figure*}
    \includegraphics[width=\textwidth]{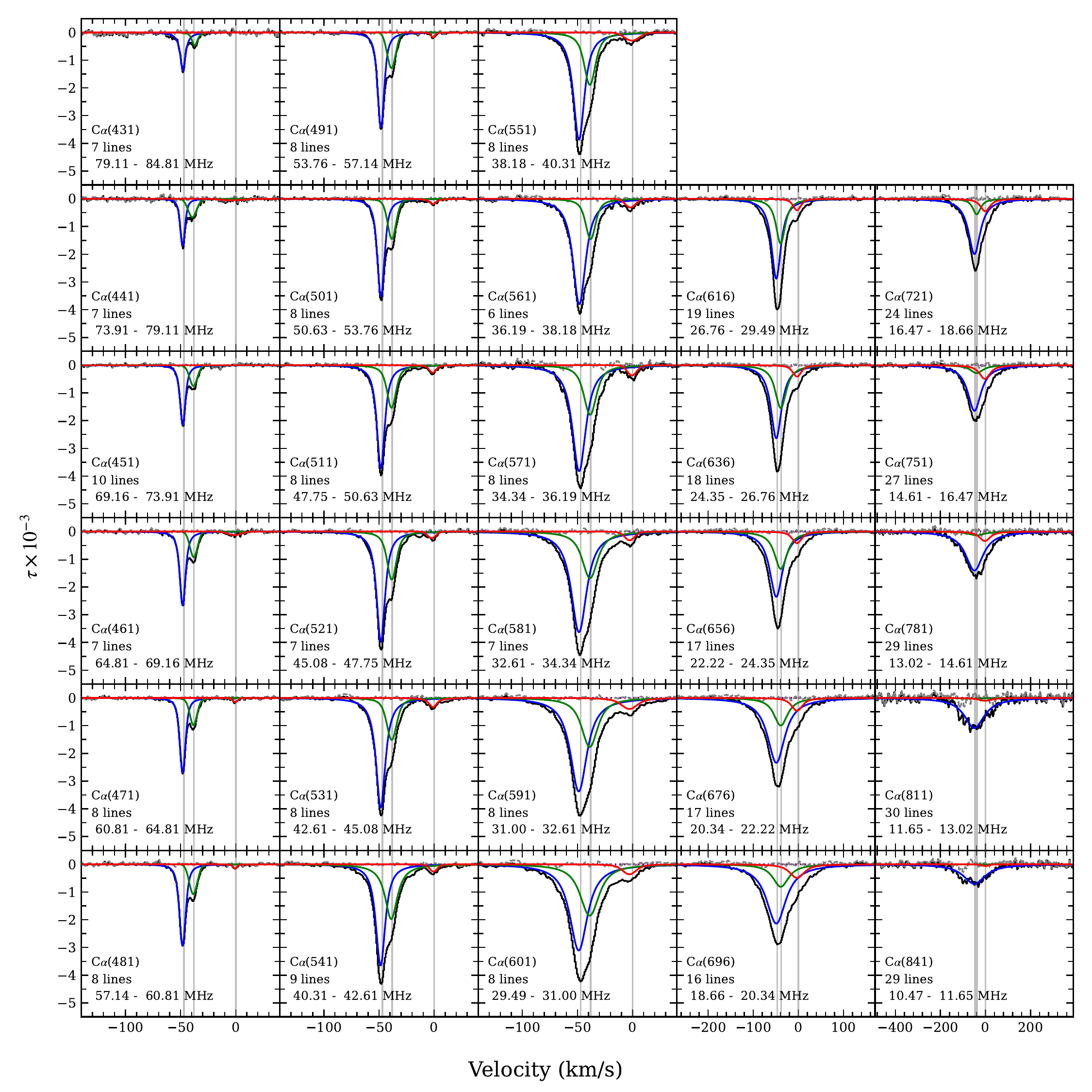}
    \caption{Every stack of lines detected towards Cassiopeia A used to extract physical parameters. The coloured lines are the results of the Voigt profile fitting algorithm. The blue lines correspond to the component at --47~km~s$^{-1}$, the green lines are the --38~km.s${-1}$ component and the red lines are the 0~km~s$^{-1}$. The measured line properties can be found in Table~\ref{tab:result-casa}.}
    \label{fig:CasA-all-detections-lines}
\end{figure*}

\begin{table*}
    \centering
    \caption{Measured line properties for C$n\alpha$ stacked recombination lines towards Cas A.}
    \begin{tabular}{|ccccc|}
        \hline
        Transition ($n$) & Frequency (MHz) & Line centroid (km~s$^{-1}$) & Line integrated intensity (m.s$^{-1}$) & Linewidth (km~s$^{-1}$)\\ \hline \hline
       431 & 81.89 & -47.95 $\pm$ 0.05 & 12.41 $\pm$ 1.56 & 5.71 $\pm$ 0.18 \\
 &  & -37.22 $\pm$ 0.2 & 3.74 $\pm$ 1.56 & 5.97 $\pm$ 2.34 \\
 &  & -2.56 $\pm$ 0.05 & 0.0 $\pm$ 1.56 & nan $\pm$ 0.0 \\ \hline
441 & 76.45 & -48.17 $\pm$ 0.04 & 22.34 $\pm$ 1.34 & 5.21 $\pm$ 0.56 \\
 &  & -39.33 $\pm$ 0.17 & 6.51 $\pm$ 1.34 & 9.19 $\pm$ 1.87 \\
 &  & -3.17 $\pm$ 1.89 & 1.34 $\pm$ 1.34 & 22.39 $\pm$ 26.67 \\ \hline
451 & 71.48 & -48.05 $\pm$ 0.02 & 23.59 $\pm$ 0.94 & 5.79 $\pm$ 0.33 \\
 &  & -38.39 $\pm$ 0.09 & 6.95 $\pm$ 0.94 & 8.2 $\pm$ 1.25 \\
 &  & -3.05 $\pm$ 158.74 & 0.03 $\pm$ 0.94 & 34.77 $\pm$ 944.41 \\ \hline
461 & 66.94 & -48.14 $\pm$ 0.02 & 27.73 $\pm$ 1.03 & 6.21 $\pm$ 0.29 \\
 &  & -38.15 $\pm$ 0.08 & 7.91 $\pm$ 1.03 & 8.03 $\pm$ 0.84 \\
 &  & -1.26 $\pm$ 0.46 & 1.97 $\pm$ 1.03 & 13.66 $\pm$ 5.74 \\ \hline
471 & 62.77 & -48.17 $\pm$ 0.02 & 27.27 $\pm$ 0.97 & 6.25 $\pm$ 0.26 \\
 &  & -38.39 $\pm$ 0.07 & 9.05 $\pm$ 0.97 & 7.72 $\pm$ 0.72 \\
 &  & -0.36 $\pm$ 0.25 & 0.89 $\pm$ 0.97 & 5.44 $\pm$ 2.72 \\ \hline
481 & 58.94 & -48.23 $\pm$ 0.02 & 26.22 $\pm$ 1.02 & 6.91 $\pm$ 0.27 \\
 &  & -38.51 $\pm$ 0.08 & 9.87 $\pm$ 1.02 & 8.59 $\pm$ 0.71 \\
 &  & -0.57 $\pm$ 0.26 & 0.95 $\pm$ 1.02 & 4.13 $\pm$ 3.31 \\ \hline
491 & 55.41 & -48.27 $\pm$ 0.02 & 31.69 $\pm$ 1.0 & 7.17 $\pm$ 0.24 \\
 &  & -38.73 $\pm$ 0.07 & 12.41 $\pm$ 1.0 & 9.05 $\pm$ 0.63 \\
 &  & -0.92 $\pm$ 0.21 & 1.1 $\pm$ 1.0 & 4.39 $\pm$ 2.57 \\ \hline
501 & 52.16 & -48.24 $\pm$ 0.02 & 37.33 $\pm$ 1.1 & 7.82 $\pm$ 0.27 \\
 &  & -38.35 $\pm$ 0.07 & 16.14 $\pm$ 1.1 & 9.11 $\pm$ 0.67 \\
 &  & -1.06 $\pm$ 0.3 & 2.17 $\pm$ 1.1 & 7.22 $\pm$ 4.25 \\ \hline
511 & 49.16 & -48.34 $\pm$ 0.03 & 47.09 $\pm$ 1.16 & 8.49 $\pm$ 0.31 \\
 &  & -38.63 $\pm$ 0.08 & 21.49 $\pm$ 1.16 & 9.96 $\pm$ 0.81 \\
 &  & -1.34 $\pm$ 0.22 & 2.58 $\pm$ 1.16 & 7.45 $\pm$ 2.55 \\ \hline
521 & 46.39 & -48.33 $\pm$ 0.03 & 56.18 $\pm$ 1.42 & 9.13 $\pm$ 0.38 \\
 &  & -38.44 $\pm$ 0.1 & 25.26 $\pm$ 1.42 & 10.27 $\pm$ 0.94 \\
 &  & -2.14 $\pm$ 0.31 & 2.25 $\pm$ 1.42 & 8.9 $\pm$ 3.56 \\ \hline
531 & 43.82 & -48.44 $\pm$ 0.05 & 62.37 $\pm$ 1.47 & 10.16 $\pm$ 0.44 \\
 &  & -38.47 $\pm$ 0.13 & 22.43 $\pm$ 1.47 & 11.01 $\pm$ 1.17 \\
 &  & -1.02 $\pm$ 0.28 & 4.51 $\pm$ 1.47 & 8.51 $\pm$ 3.62 \\ \hline
541 & 41.44 & -48.61 $\pm$ 0.04 & 58.27 $\pm$ 1.31 & 10.14 $\pm$ 0.45 \\
 &  & -38.93 $\pm$ 0.13 & 41.92 $\pm$ 1.31 & 13.53 $\pm$ 1.03 \\
 &  & -0.55 $\pm$ 0.28 & 2.56 $\pm$ 1.31 & 8.21 $\pm$ 3.09 \\ \hline
551 & 39.22 & -48.61 $\pm$ 0.08 & 68.39 $\pm$ 1.6 & 11.99 $\pm$ 0.6 \\
 &  & -38.73 $\pm$ 0.19 & 34.13 $\pm$ 1.6 & 12.54 $\pm$ 1.46 \\
 &  & -0.56 $\pm$ 0.43 & 5.03 $\pm$ 1.6 & 15.96 $\pm$ 5.03 \\ \hline
561 & 37.17 & -48.27 $\pm$ 0.12 & 77.02 $\pm$ 2.08 & 13.79 $\pm$ 0.84 \\
 &  & -38.04 $\pm$ 0.26 & 22.5 $\pm$ 2.08 & 11.02 $\pm$ 2.0 \\
 &  & -1.98 $\pm$ 0.49 & 3.71 $\pm$ 2.08 & 11.16 $\pm$ 4.69 \\ \hline
571 & 35.25 & -48.43 $\pm$ 0.27 & 85.71 $\pm$ 2.64 & 14.75 $\pm$ 1.33 \\
 &  & -38.43 $\pm$ 0.55 & 35.89 $\pm$ 2.64 & 13.93 $\pm$ 2.93 \\
 &  & -0.43 $\pm$ 0.71 & 4.1 $\pm$ 2.64 & 10.26 $\pm$ 4.98 \\ \hline
581 & 33.46 & -48.6 $\pm$ 0.35 & 87.82 $\pm$ 1.89 & 16.76 $\pm$ 1.3 \\
 &  & -38.6 $\pm$ 0.79 & 42.54 $\pm$ 1.89 & 17.0 $\pm$ 3.44 \\
 &  & -2.6 $\pm$ 0.8 & 4.1 $\pm$ 1.89 & 10.83 $\pm$ 5.13 \\ \hline
591 & 31.79 & -48.82 $\pm$ 0.46 & 88.87 $\pm$ 1.76 & 18.1 $\pm$ 1.53 \\
 &  & -38.82 $\pm$ 0.87 & 46.81 $\pm$ 1.76 & 17.75 $\pm$ 3.67 \\
 &  & -2.78 $\pm$ 0.99 & 10.35 $\pm$ 1.76 & 18.09 $\pm$ 5.97 \\ \hline
601 & 30.23 & -48.83 $\pm$ 0.81 & 84.87 $\pm$ 2.0 & 19.43 $\pm$ 2.35 \\
 &  & -38.88 $\pm$ 1.74 & 57.67 $\pm$ 2.0 & 20.79 $\pm$ 5.84 \\
 &  & -2.83 $\pm$ 1.63 & 7.38 $\pm$ 2.0 & 15.24 $\pm$ 6.98 \\ \hline
616 & 28.08 & -49.0 $\pm$ 2.21 & 105.32 $\pm$ 1.43 & 25.34 $\pm$ 4.28 \\
 &  & -39.38 $\pm$ 4.39 & 59.98 $\pm$ 1.43 & 24.74 $\pm$ 8.84 \\
 &  & -3.0 $\pm$ 4.3 & 12.03 $\pm$ 1.43 & 20.64 $\pm$ 5.57 \\ \hline
 636 & 25.51 & -49.0 $\pm$ 3.2 & 108.48 $\pm$ 56.3 & 27.74 $\pm$ 5.6 \\
 &  & -39.0 $\pm$ 7.89 & 75.62 $\pm$ 52.83 & 28.38 $\pm$ 3.34 \\ 
 &  & -3.0 $\pm$ 6.29 & 15.96 $\pm$ 14.1 & 23.45 $\pm$ 12.99 \\ \hline
    \end{tabular}
    \tablefoot{For $n\leq 636$, the uncertainties on the measurements are computed with the correlation matrix of the least-square optimisation.}
    \label{tab:result-casa-0}
\end{table*}

\begin{table*}
    \centering
    \caption{Measured line properties for C$n\alpha$ stacked recombination lines towards Cas A.}
    \begin{tabular}{|ccccc|}
        \hline
        Transition ($n$) & Frequency (MHz) & Line centroid (km~s$^{-1}$) & Line integrated intensity (m.s$^{-1}$) & Linewidth (km~s$^{-1}$)\\ \hline \hline
656 & 23.25 & -50.67 $\pm$ 8.94 & 101.45 $\pm$ 49.1 & 32.99 $\pm$ 4.47 \\
 &  & -41.83 $\pm$ 10.57 & 85.82 $\pm$ 41.84 & 31.07 $\pm$ 3.09 \\
 &  & -3.45 $\pm$ 9.34 & 21.46 $\pm$ 14.59 & 28.29 $\pm$ 11.23 \\ \hline
676 & 21.25 & -51.11 $\pm$ 6.67 & 138.98 $\pm$ 26.74 & 44.17 $\pm$ 7.58 \\
 &  & -42.31 $\pm$ 7.86 & 73.46 $\pm$ 23.71 & 33.9 $\pm$ 5.77 \\
 &  & -3.77 $\pm$ 7.53 & 28.84 $\pm$ 24.02 & 32.94 $\pm$ 18.06 \\ \hline
696 & 19.47 & -50.88 $\pm$ 6.67 & 153.01 $\pm$ 36.94 & 52.37 $\pm$ 13.71 \\
 &  & -41.42 $\pm$ 7.51 & 72.79 $\pm$ 18.21 & 40.74 $\pm$ 10.47 \\
 &  & -4.24 $\pm$ 7.47 & 33.55 $\pm$ 20.63 & 39.72 $\pm$ 14.13 \\ \hline
721 & 17.52 & -49.99 $\pm$ 2.69 & 159.67 $\pm$ 29.36 & 63.07 $\pm$ 15.3 \\
 &  & -40.01 $\pm$ 2.59 & 76.7 $\pm$ 14.87 & 48.9 $\pm$ 11.27 \\
 &  & -2.6 $\pm$ 4.01 & 29.01 $\pm$ 21.99 & 45.74 $\pm$ 25.3 \\ \hline
751 & 15.5 & -50.25 $\pm$ 4.58 & 143.21 $\pm$ 47.5 & 69.09 $\pm$ 14.4 \\
 &  & -40.4 $\pm$ 4.47 & 70.41 $\pm$ 25.01 & 68.84 $\pm$ 20.52 \\
 &  & -3.83 $\pm$ 5.26 & 38.83 $\pm$ 40.62 & 55.54 $\pm$ 39.94 \\ \hline
781 & 13.78 & -48.84 $\pm$ 1.97 & 147.56 $\pm$ 43.05 & 94.67 $\pm$ 23.13 \\
 &  & -38.85 $\pm$ 1.99 & 73.69 $\pm$ 27.26 & 88.45 $\pm$ 30.04 \\
 &  & -2.66 $\pm$ 2.51 & 30.5 $\pm$ 47.66 & 55.49 $\pm$ 49.11 \\ \hline
811 & 12.31 & -47.73 $\pm$ 0.87 & 115.67 $\pm$ 52.17 & 100.45 $\pm$ 26.15 \\
 &  & -37.73 $\pm$ 0.87 & 56.84 $\pm$ 25.79 & 111.69 $\pm$ 51.63 \\
 &  & -1.72 $\pm$ 0.85 & 16.09 $\pm$ 51.42 & 75.15 $\pm$ 87.73 \\ \hline
841 & 11.04 & -49.74 $\pm$ 3.06 & 84.39 $\pm$ 59.75 & 120.99 $\pm$ 131.2 \\
 &  & -39.74 $\pm$ 3.06 & 42.07 $\pm$ 28.16 & 119.73 $\pm$ 74.17 \\
 &  & -3.56 $\pm$ 2.93 & 4.72 $\pm$ 20.25 & 70.71 $\pm$ 148.74 \\ \hline
    \end{tabular}
    \tablefoot{For $n>636$, the uncertainties on the measurements are computed based on the statistical method described in Sect.~\ref{subsub:casao-linefitting}.}
    \label{tab:result-casa-1}
\end{table*}

\begin{figure*}
    \includegraphics[width=\textwidth]{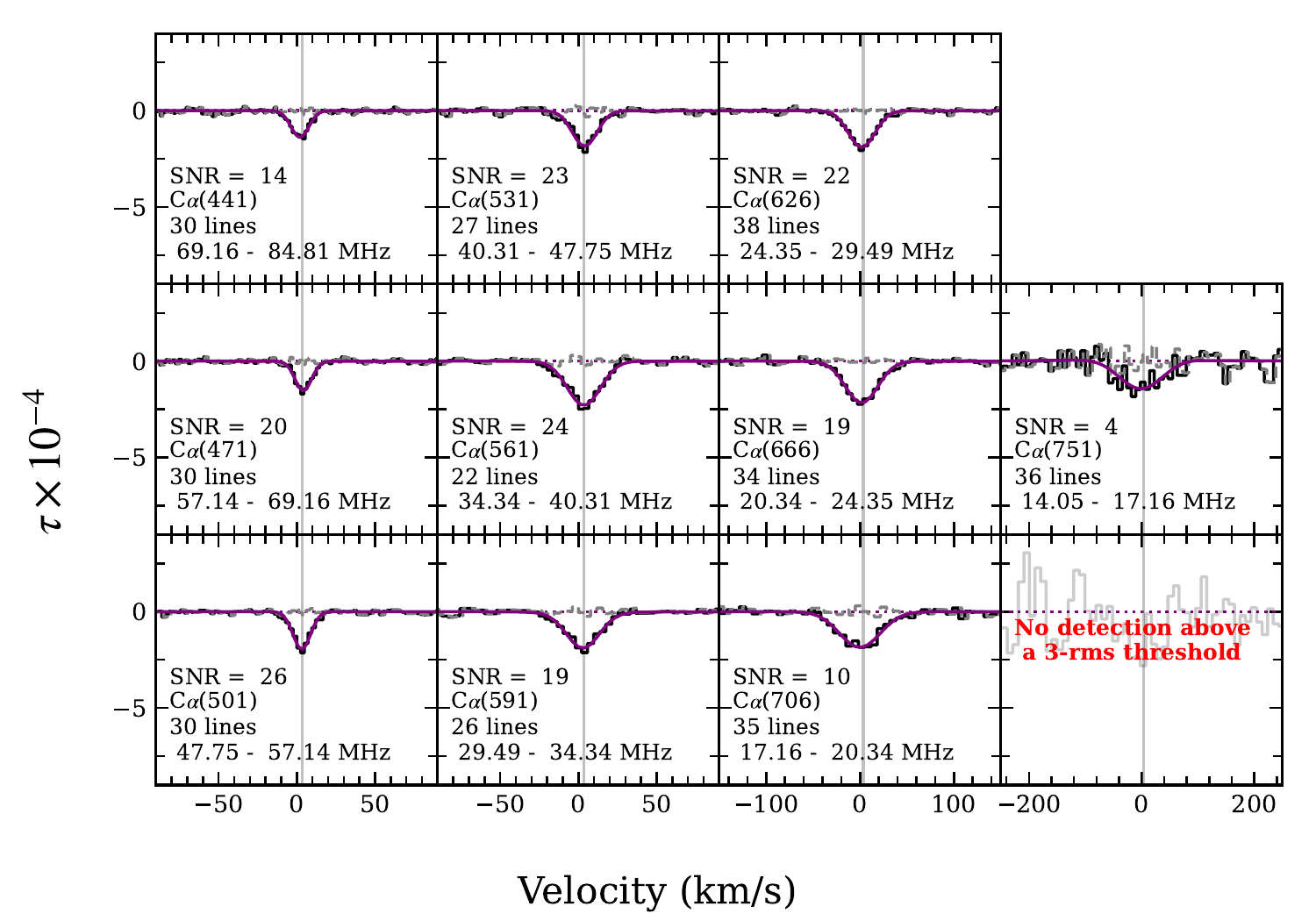}
    \caption{Every stack of lines detected towards Cygnus A used to extract physical parameters. The purple lines are the results of the Voigt profile fitting algorithm.}
    \label{fig:CygA-detections-lines}
\end{figure*}

\begin{table*}
    \centering
    \caption{Measured line properties for C$n\alpha$ stacked recombination lines towards Cyg A.}
    \begin{tabular}{|ccccc|}
        \hline
        Transition ($n$) & Frequency (MHz) & Line centroid (km~s$^{-1}$) & Line integrated intensity (m~s$^{-1}$) & Linewidth (km~s$^{-1}$)\\ \hline \hline
        441 & 76.45 &  1.98 $\pm$ 0.24 & 1.97 $\pm$ 0.59 & 13.43 $\pm$ 2.87 \\
        471 & 62.77 &  4.15 $\pm$ 0.19 & 2.04 $\pm$ 0.53 & 12.46 $\pm$ 2.23 \\
        501 & 52.16 &  3.15 $\pm$ 0.16 & 2.74 $\pm$ 0.57 & 13.18 $\pm$ 1.92 \\
        531 & 43.82 &  3.98 $\pm$ 0.27 & 3.67 $\pm$ 0.74 & 18.73 $\pm$ 3.15 \\
        561 & 37.17 &  3.5 $\pm$ 0.3   & 6.07 $\pm$ 0.87 & 24.92 $\pm$ 3.46 \\
        591 & 31.79 &  2.86 $\pm$ 0.41 & 5.27 $\pm$ 0.99 & 25.09 $\pm$ 4.92 \\
        626 & 26.76 &  1.78 $\pm$ 0.45 & 6.77 $\pm$ 0.94 & 33.47 $\pm$ 5.22 \\
        666 & 22.22 &  1.72 $\pm$ 0.58 & 9.36 $\pm$ 1.26 & 40.69 $\pm$ 6.77 \\
        706 & 18.66 &  -0.19 $\pm$ 1.26 & 10.05 $\pm$ 2.1 & 50.94 $\pm$ 14.81 \\
        751 & 15.50 &  -1.21 $\pm$ 6.42 & 11.41 $\pm$ 6.62 & 85.25 $\pm$ 72.34 \\ \hline
    \end{tabular}
    \tablefoot{The uncertainties on the measurements are computed with the correlation matrix of the least-square optimisation.}
    \label{tab:result-cyga}
\end{table*}

\FloatBarrier
\section{Result of the grid search}

\begin{figure*}[h!]
    \centering
    \includegraphics[width=\textwidth]{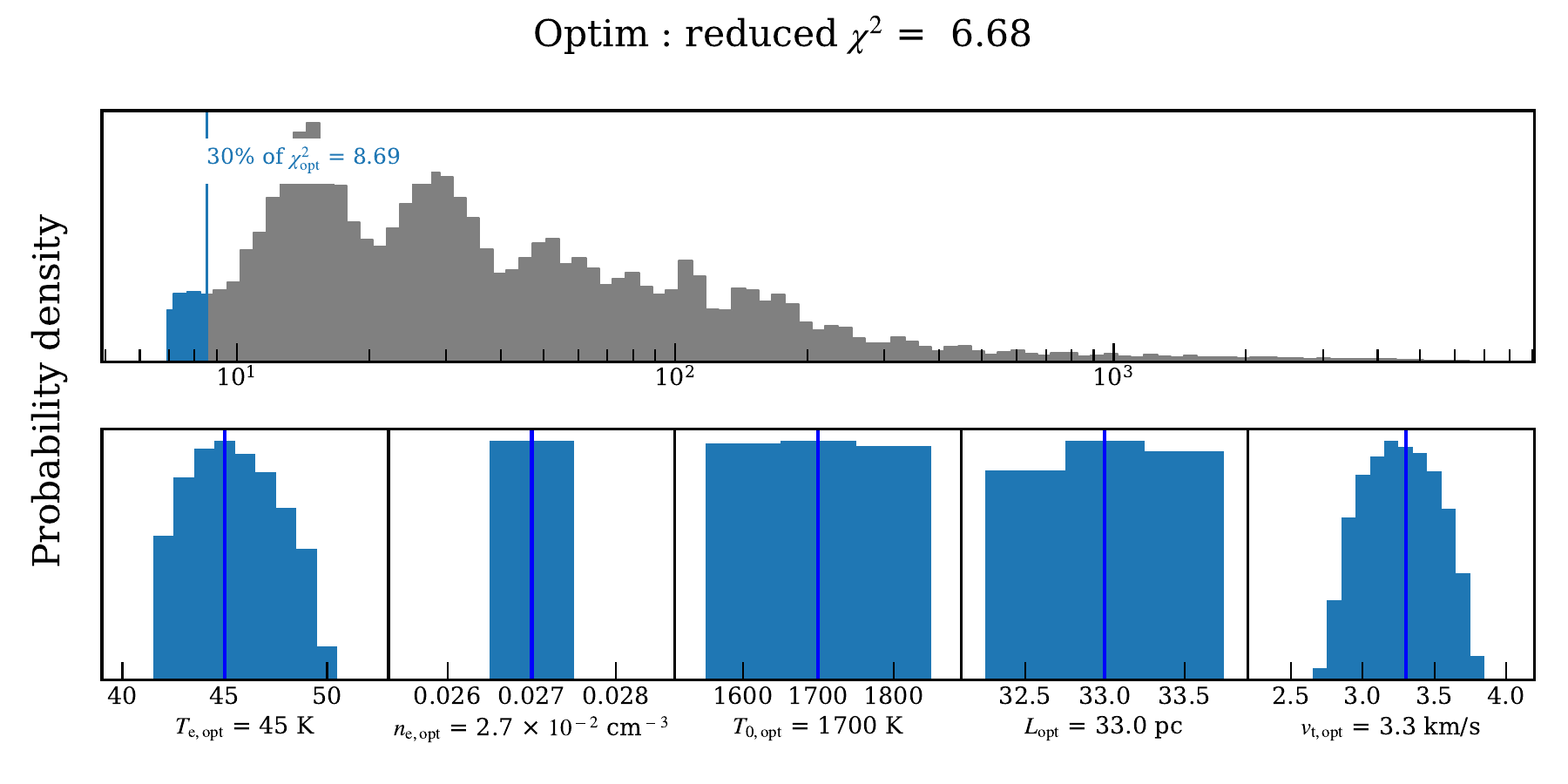}
    \caption{Summary of the grid search for Cassiopeia A, component -47 ~km~s$^{-1}$. The top panel is the histogram of all the nodes of the final grid. The blue part of the histogram highlights the 30\% of $\chi^2$ that are kept to evaluate the uncertainties. The bottom panels represent for each parameter the values that gives a $\chi^2$ distance comprised in the 30\% range.}
    \label{fig:casa-47-histo}
\end{figure*}

\begin{figure*}
    \centering
    \includegraphics[width=\textwidth]{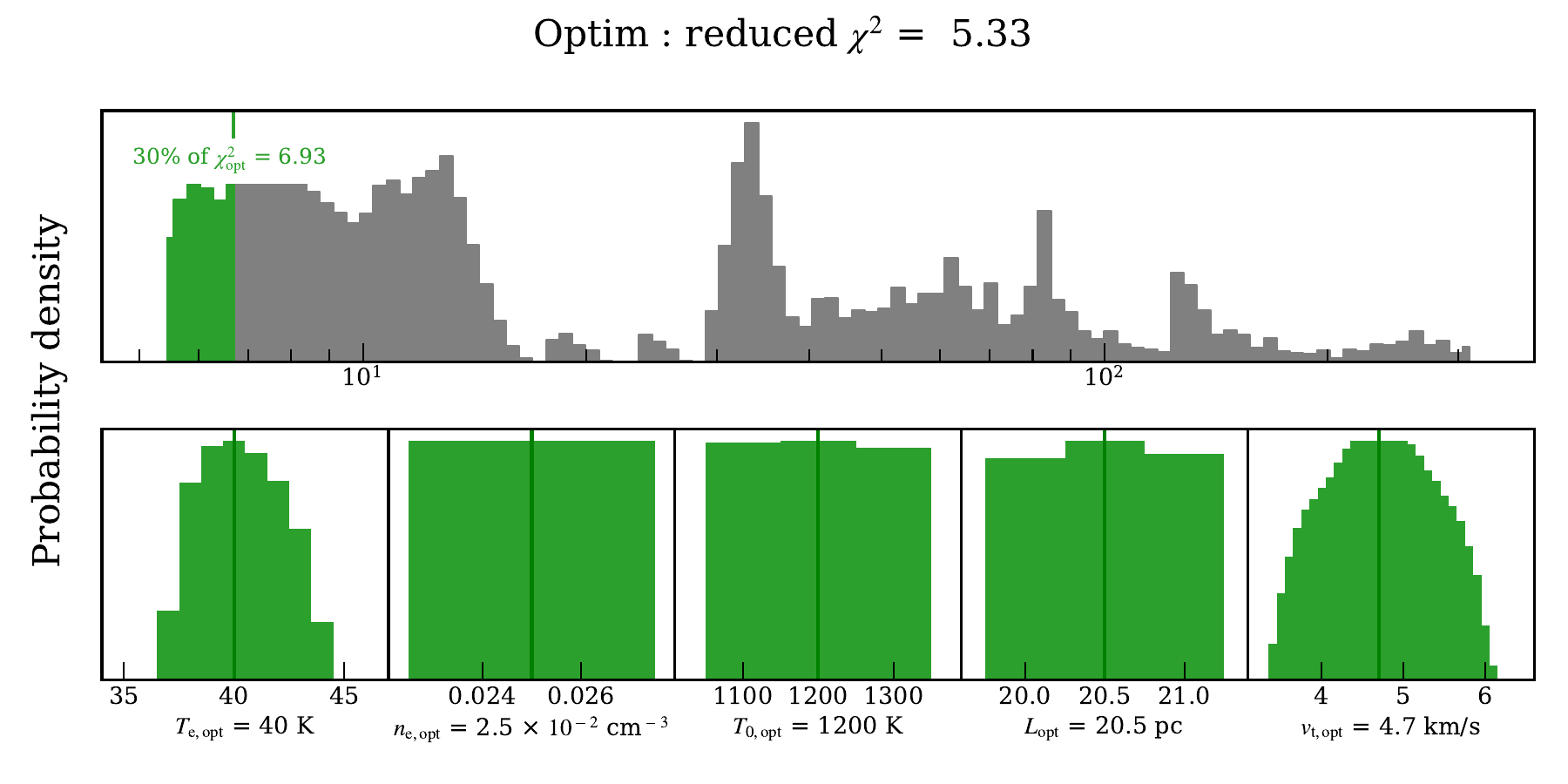}
    \caption{Summary of the grid search for Cassiopeia A, component -38 ~km~s$^{-1}$. The green part of the histogram highlights the 30\% of $\chi^2$ that are kept to evaluate the uncertainties. The bottom panels represent for each parameter the values that gives a $\chi^2$ distance comprised in the 30\% range.}
    \label{fig:casa-38-histo}
\end{figure*}

\begin{figure*}
    \centering
    \includegraphics[width=\textwidth]{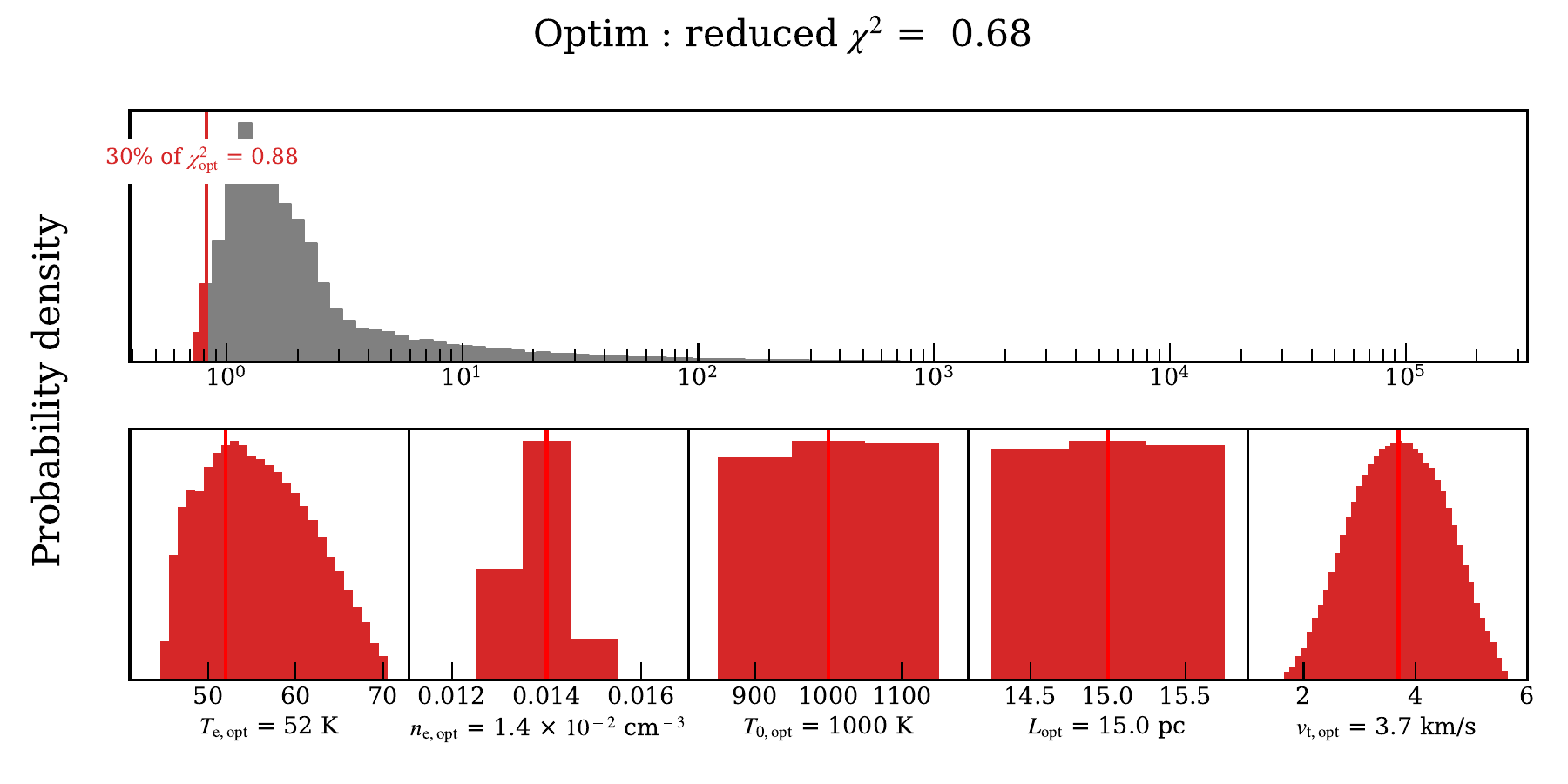}
    \caption{Summary of the grid search for Cassiopeia A, component -47 ~km~s$^{-1}$. The red part of the histogram highlights the 30\% of $\chi^2$ that are kept to evaluate the uncertainties. The bottom panels represent for each parameter the values that gives a $\chi^2$ distance comprised in the 30\% range.}
    \label{fig:casa-0-histo}
\end{figure*}
\begin{figure*}
    \centering
    \includegraphics[width=\textwidth]{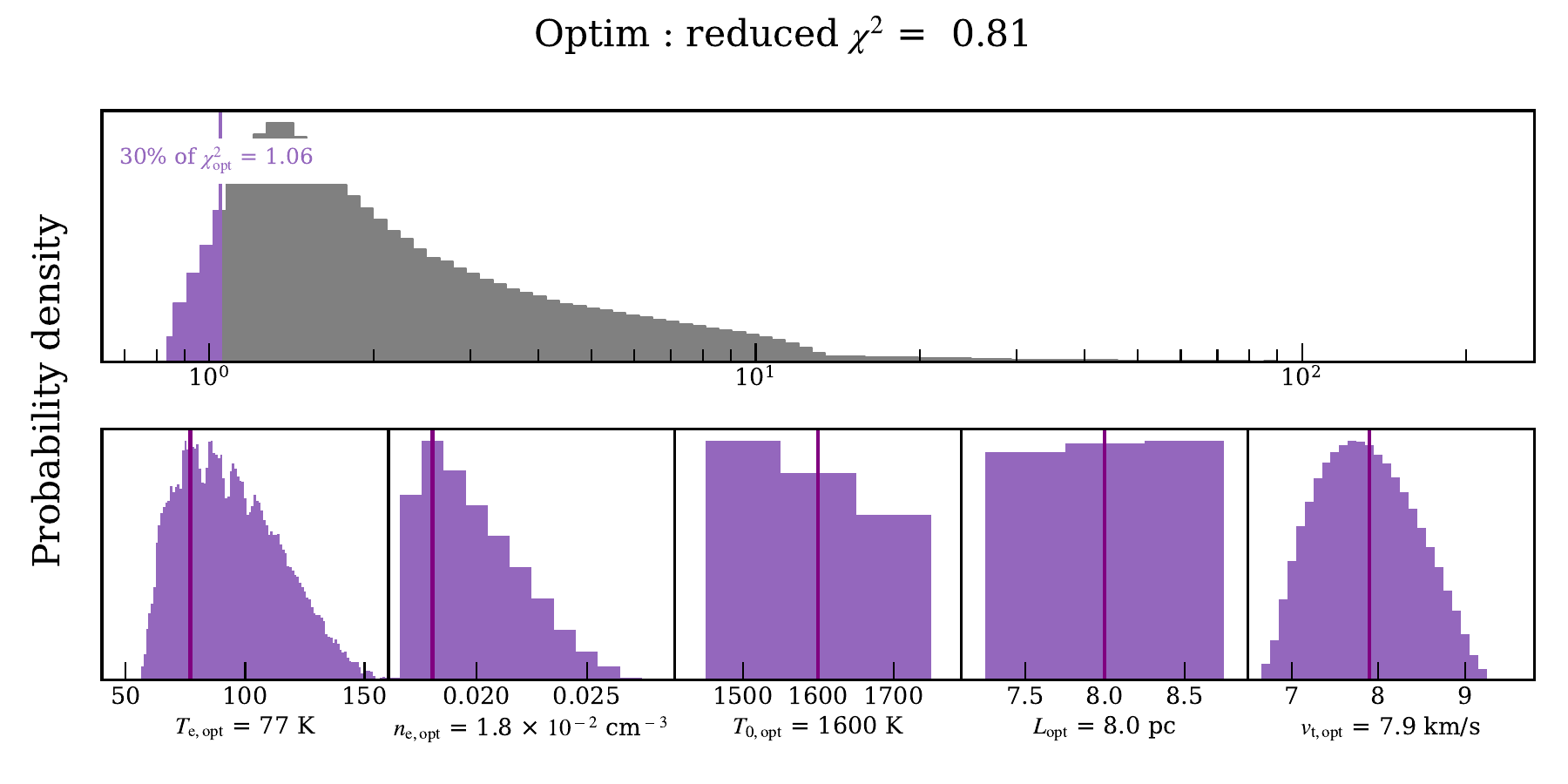}
    \caption{Summary of the grid search for Cygnus A, component 3.5 ~km~s$^{-1}$. The purple part of the histogram highlights the 30\% of $\chi^2$ that are kept to evaluate the uncertainties. The bottom panels represent for each parameter the values that gives a $\chi^2$ distance comprised in the 30\% range.}
    \label{fig:cyga-histo}
\end{figure*}

\end{appendix}

\end{document}